\documentclass[manuscript]{aastex}

\def\lax{{$\mathrel{\hbox{\rlap{\hbox{\lower4pt\hbox{$\sim$}}}\hbox{$<$}}}$}}
\def\gax{{$\mathrel{\hbox{\rlap{\hbox{\lower4pt\hbox{$\sim$}}}\hbox{$>$}}}$}}

\usepackage{color}

\usepackage{lscape}

\slugcomment{}

\begin{document}
\title{Classification of Tidal Disruption Events Based on Stellar Orbital Properties}

 \author{
Kimitake \textsc{Hayasaki}\altaffilmark{1},
Shiyan \textsc{Zhong}\altaffilmark{2},
Shuo \textsc{Li}\altaffilmark{3},
Peter \textsc{Berczik} \altaffilmark{3,4,5},
and
Rainer \textsc{Spurzem}\altaffilmark{3,4,6}
}
\altaffiltext{1}{Department of Astronomy and Space Science, Chungbuk National University, Cheongju 361-763, Korea}
\email{kimi@cbnu.ac.kr}
\altaffiltext{2}{Yunnan Observatories, Chinese Academy of Sciences, 396 Yang-Fang-Wang, Guandu District , 650216, Kunming, Yunnan, China}
\altaffiltext{3}{National Astronomical Observatories of China and Key Laboratory for Computational Astrophysics, Chinese Academy of Sciences, 20A Datun Rd., Chaoyang District, 100012, Beijing, China}
\altaffiltext{4}{Astronomisches Rechen-Institut, Zentrum f$\ddot{\rm u}$r Astronomie, 
University of Heidelberg, M$\ddot{\rm o}$nchhofstrasse 12-14, D-69120 Heidelberg, Germany}
\altaffiltext{5}{Main Astronomical Observatory, National Academy of Sciences of Ukraine, 27 Akademika Zabolotnoho St., 03680 Kyiv, Ukraine}
\altaffiltext{6}{Kavli Institute for Astronomy and Astrophysics, Peking University, 100871 Beijing, China}
%
\begin{abstract}
We study the rates of tidal disruption of stars by intermediate-mass 
to supermassive black holes on bound to unbound orbits by using 
high-accuracy direct N-body experiments. The approaching stars from 
the star cluster to the black hole can take three types of orbit: eccentric, 
parabolic, and hyperbolic orbits. Since the mass fallback rate shows a 
different variability depending on these orbital types, we can classify tidal 
disruption events (TDEs) into three main categories: eccentric, parabolic, 
and hyperbolic TDEs.\,\,Respective TDEs are characterized by two critical 
values of the orbital eccentricity: the lower critical eccentricity is the one 
below which the stars on eccentric orbits cause the finite, intense accretion, 
and the higher critical eccentricity above which the stars on hyperbolic 
orbits cause no accretion. Moreover, we find that the parabolic TDEs 
are divided into three subclasses: precisely parabolic, marginally eccentric, 
and marginally hyperbolic TDEs. We analytically derive that the 
mass fallback rate of the marginally eccentric TDEs can be flatter and slightly 
higher than the standard fallback rate proportional to $t^{-5/3}$, whereas it 
can be flatter and lower for the marginally hyperbolic TDEs. We confirm by 
N-body experiments that only few eccentric, precisely parabolic, and hyperbolic 
TDEs can occur in a spherical stellar system with a single intermediate-mass 
to supermassive black hole. A substantial fraction of the stars approaching to 
the black hole would cause the marginally eccentric or marginally hyperbolic TDEs.
\end{abstract}
%
\section{Introduction}
%

%
%
Tidal disruption events (TDEs) are thought to be a key phenomenon 
for searching dormant supermassive black holes (SMBHs) at the centers 
of the inactive galaxies or intermediate-mass black holes (IMBHs) at those 
of the star clusters. Most TDEs take place when a star at large separation 
($\sim1\,\rm{pc}$) is perturbed onto a parabolic orbit approaching close 
enough to the SMBH to be ripped apart by its tidal force. The subsequent 
accretion of stellar debris falling back to the SMBH causes a characteristic 
flare with a luminosity large enough to exceed the Eddington luminosity for 
a time scale of weeks to months \citep{rees88,p89,ek89}. Such flares have 
been discovered at optical \citep{sg+12,ar14,ho14,2017ApJ...842...29H}, 
ultraviolet \citep{2006ApJ...653L..25G,2014ApJ...780...44C,2015ApJ...798...12V}, and 
soft X-ray (\citealt{kb99}; \citealt{2012A&A...541A.106S}; \citealt{2013MNRAS.435.1904M}; 
\citealt{2017ApJ...838..149A}) wavelengths with inferred event rates of 
$10^{-4}-10^{-5}$ per year per galaxy \citep{dbeb02, jd04,vf14,sm16}. 
The other, high-energy jetted TDEs have been detected through non-thermal 
emissions in radio \citep{baz+11,2016ApJ...819L..25A,2016Sci...351...62V} or hard X-ray 
\citep{dnb+11,2015MNRAS.452.4297B} wavelengths with much lower event rate 
\citep{2014arXiv1411.0704F}.

%
%
TDEs can largely contribute to the growth of the relatively low-mass 
SMBHs ($\la10^{6}M_\odot$) or IMBHs because of the lack of large 
amount of gas in their environments, although the rate of the tidal 
disruption is relatively low. The growth rate depends on the stellar 
density profile \citep{bw76} and timescale of mass supply in the star 
cluster based on the classical loss cone theory \citep{fr76}. \cite{bme04} 
examined the cluster density profile and the effect of the TDEs on the 
black hole growth by performing the self-consistent N-body simulations 
of star clusters composed of equal-mass stars and a central, IMBH. 
Subsequently, \cite{bbk11} calculated the tidal disruption rate of stars 
by SMBHs by performing higher resolution N-body simulations. They 
concluded that relaxation-driven stellar feeding cannot let the black hole 
grow to more than $10^{7}\,M_{\odot}$. Although the standard two-body 
scattering mechanism for generating TDEs \citep{mt99,jd04} predicts effectively 
parabolic trajectories, recent high-accuracy direct N-body simulations show that 
a significant amount of stars entering the tidal disruption radius has the orbital 
eccentricities less or more than $1.0$ \citep{zhong+14}. 

%
%
It still remains under debate how the standard, theoretical mass fallback rate 
proportional to $t^{^{-5/3}}$ \citep{rees88,p89,ek89} translates into the observed 
light curves. While most of the soft X-ray TDEs appear to follow the $t^{{-5/3}}$ power law 
decay curve proportional to the fallback rate (see \citealt{2015JHEAp...7..148K} for a review), 
the optical to ultraviolet TDEs exhibit the different decay curve 
\citep{sg+12,2014ApJ...780...44C,ar14,ho14}.

%
%
\cite{lkp09} numerically showed that the fallback rate depends on the internal structure 
of the tidally disrupted stars, leading to early-time deviations from the standard fallback 
rate. The centrally condensed core survived by the partial disruption of the star can let
the resultant light curves steeper \citep{2013ApJ...767...25G}.
The accretion of clumps formed by the self-gravity of the debris stream 
causes the significant variations of the light curve around the $t^{-5/3}$ 
average at late times \citep{2015ApJ...808L..11C}. The outflows or winds 
caused during the super-Eddington accretion phase lets the optical to 
ultraviolet light curves deviated from the standard $t^{-5/3}$ curve 
\citep{2009MNRAS.400.2070S,lr11}. There have been some arguments 
that the energy dissipated by the stream-stream collisions during the debris 
circularization powers the observed optical to ultraviolet TDEs 
\citep{2015ApJ...806..164P,2016ApJ...830..125J,2017MNRAS.464.2816B}.

%
%
Recent hydrodynamic simulations have shown that observable properties of these 
``eccentric'' TDEs significantly deviate from those of standard TDEs; in particular, 
the rate of mass return is substantially increased by being cut off at a finite time, 
rather than continuing indefinitely as a power law decay \citep{hsl13,hsl16}. 
This suggests that the variability of TDE light curves also depends on the orbital type 
of approaching stars, especially orbital eccentricity and penetration factor (which is 
the ratio of the tidal disruption radius to pericenter distance of the star) of stars.

%
%
In this paper, we classify the TDEs by the type of orbits of stars approaching to 
SMBHs or IMBHs, and examine each occurrence rate in the dense star cluster 
system modeled by N-body experiments. In Section~\ref{sec:2}, we give a condition 
to classify the TDEs by the type of the stellar orbits, and analytically derive the mass fallback 
rate of each TDE based on the condition, which can have the different time dependence 
from the standard fallback rate proportional to $t^{^-5/3}$. In Section~\ref{sec:3}, we 
describe our numerical approach and simulations results, where we mainly focus on 
the eccentricity distribution of the N-body particles for their penetration factor. We 
discuss the reality of our simulation results by using the scaling method to extrapolate 
them in Section~\ref{sec:dis}. Finally, Section~\ref{sec:con} is devoted to conclusion of 
our scenario.

%
\section{Type of tidal disruption events}
\label{sec:2}
%

As a star approaches and enters into the tidal disruption radius 
of the SMBH or IMBH, it is disrupted by the tidal force of the black hole 
which dominating the stellar self-gravity and pressure forces at 
the tidal disruption radius:
\begin{equation}
 r_{t}=\left(\frac{M_{\rm bh}}{m_*}\right)^{1/3}r_{*}\approx
24\left(\frac{M_{\rm bh}}{10^6\,{M}_\odot}\right)^{-2/3}
\left(\frac{m_*}{{M}_\odot}\right)^{-1/3}
\left(\frac{r_*}{{R}_\odot}\right)
r_{\rm S}.
\label{eq:rt}
\end{equation} 
Here we denote the black hole mass with $M_{\rm bh}$, stellar mass 
with $m_*$ and radius with $r_*$, and the Schwarzschild radius 
with $r_{\rm{S}}=2{\rm G}M_{\rm bh}/{\rm c}^2$, where $G$ and 
$c$ are Newton's gravitational constant and the speed of light, 
respectively. The tidal force then produces a spread in specific 
energy of the stellar debris:
\begin{equation}
\Delta\epsilon\approx \frac{GM_{\rm bh}r_*}{r_{\rm t}^2}
\label{eq:spreade}
\end{equation}
\citep{ek89}.
 
%
\subsection{Critical value of orbital eccentricity and semi-major axis}
%
The specific energy of the tidally disrupted star ranges over
\begin{eqnarray}
-\Delta\epsilon+\epsilon_{\rm orb}
\le\epsilon\le
\Delta\epsilon+\epsilon_{\rm orb}.
\label{eq:enedis}
\end{eqnarray}
Here $\epsilon_{\rm orb}$ is the specific orbital energy of the star approaching to the black hole:
\begin{eqnarray}
\epsilon_{\rm orb}=\left\{ \begin{array}{ll}
-\frac{GM_{\rm bh}}{2r_{\rm{t}}}\beta(1-e) & {\rm eccentric \,or\, circular \,\,orbit}:(0\le{e}<1) \\
0 & {\rm parabolic \,\,orbit}:(e=1) \\
\frac{GM_{\rm bh}}{2r_{\rm{t}}}\beta(e-1) & {\rm hyperbolic \,\,orbit}:(e>1), \\
\end{array} \right.
\end{eqnarray}
where $e$ and $\beta$ are the orbital eccentricity of 
the approaching star and the penetration factor, respectively. 
The penetration factor is defined by $r_{\rm t}/r_{\rm p}$, where $r_{\rm p}$ is 
the pericenter distance: $r_{\rm p}=a(1-e)$ for eccentric orbits 
and $r_{\rm p}=a(e-1)$ for hyperbolic orbits. 
In the standard TDE scenario that a star is disrupted from a parabolic orbit, 
the debris mass will be centered on zero and distributed 
over $-\Delta \epsilon\le\epsilon\le\Delta\epsilon$ because of $\epsilon_{\rm orb}=0$
\citep{rees88, ek89}.

Since the stellar debris with negative specific energy 
is bound to the black hole, it returns to pericenter and 
will eventually accrete onto the black hole. 
For eccentric orbits, if $\Delta\epsilon+\epsilon_{\rm orb}\le0$ in equation~(\ref{eq:enedis}), 
all the stellar debris should be bounded by the black hole even after the tidal disruption, 
and eventually fallbacks to the black hole. The condition $\epsilon_{\rm{orb}}=-\Delta\epsilon$ 
therefore gives a critical value of orbital eccentricity of the star
\begin{equation}
e_{\rm crit,1}=1-2\frac{q^{-1/3}}{\beta},
\label{eq:ec1}
\end{equation}
below which all the stellar debris should remain gravitationally 
bound to the black hole, where the ratio of the black hole to stellar mass 
is defined by $q\equiv{M}_{\rm bh}/m_{*}$.

In contrast, if $-\Delta\epsilon+\epsilon_{\rm orb}\le0$ in equation~(\ref{eq:enedis}) 
for the hyperbolic orbits, a part of the stellar debris should be bounded by the black 
hole and eventually fallbacks to the black hole. The condition $\epsilon_{\rm{orb}}=\Delta\epsilon$ 
also gives a critical value of orbital eccentricity of the star
\begin{equation}
e_{\rm crit,2}=1+2\frac{q^{-1/3}}{\beta}
\label{eq:ec2}
\end{equation}
below which a part of the stellar debris should remain gravitationally bound to the black hole.

These critical eccentricities give us the condition that the tidal disruption flare can happen 
in terms of the orbital eccentricity of the star:
\begin{eqnarray}
\left\{ \begin{array}{ll}
0 \le e < e_{\rm crit,1} & {\rm eccentric \,\, TDEs} \\
e_{\rm crit,1} \le e \le e_{\rm crit,2} & {\rm parabolic \,\,TDEs} \\
e_{\rm crit,2} < e & {\rm hyperbolic \,\,TDEs}. \\
\end{array} \right.
\label{eq:eclass}
\end{eqnarray}
Alternatively, we can define the critical value to classify the TDEs 
from the viewpoints of the semi-major axis as follows:
\begin{eqnarray}
\left\{ \begin{array}{ll}
0 < a < a_{\rm c} & {\rm eccentric \,\, TDEs} \\
 a_{\rm c} \le a & {\rm parabolic \,\,TDEs} \\
0<a<{a}_{\rm c} & {\rm hyperbolic \,\,TDEs}, \\
\end{array} \right.
\label{eq:aclass}
\end{eqnarray}
where {$\epsilon_{\rm orb}<0$ for the eccentric TDEs and $\epsilon_{\rm orb}>0$ 
for the hyperbolic TDEs, and} $a_{\rm c}$ is defined by
\begin{eqnarray}
a_{\rm c}\equiv\frac{q^{1/3}}{2}r_{\rm t}=50\left(\frac{q}{10^6}\right)^{1/3}\,r_{\rm t}.
\label{eq:ac}
\end{eqnarray}

\begin{figure}[!ht]
\resizebox{\hsize}{!}{
\includegraphics{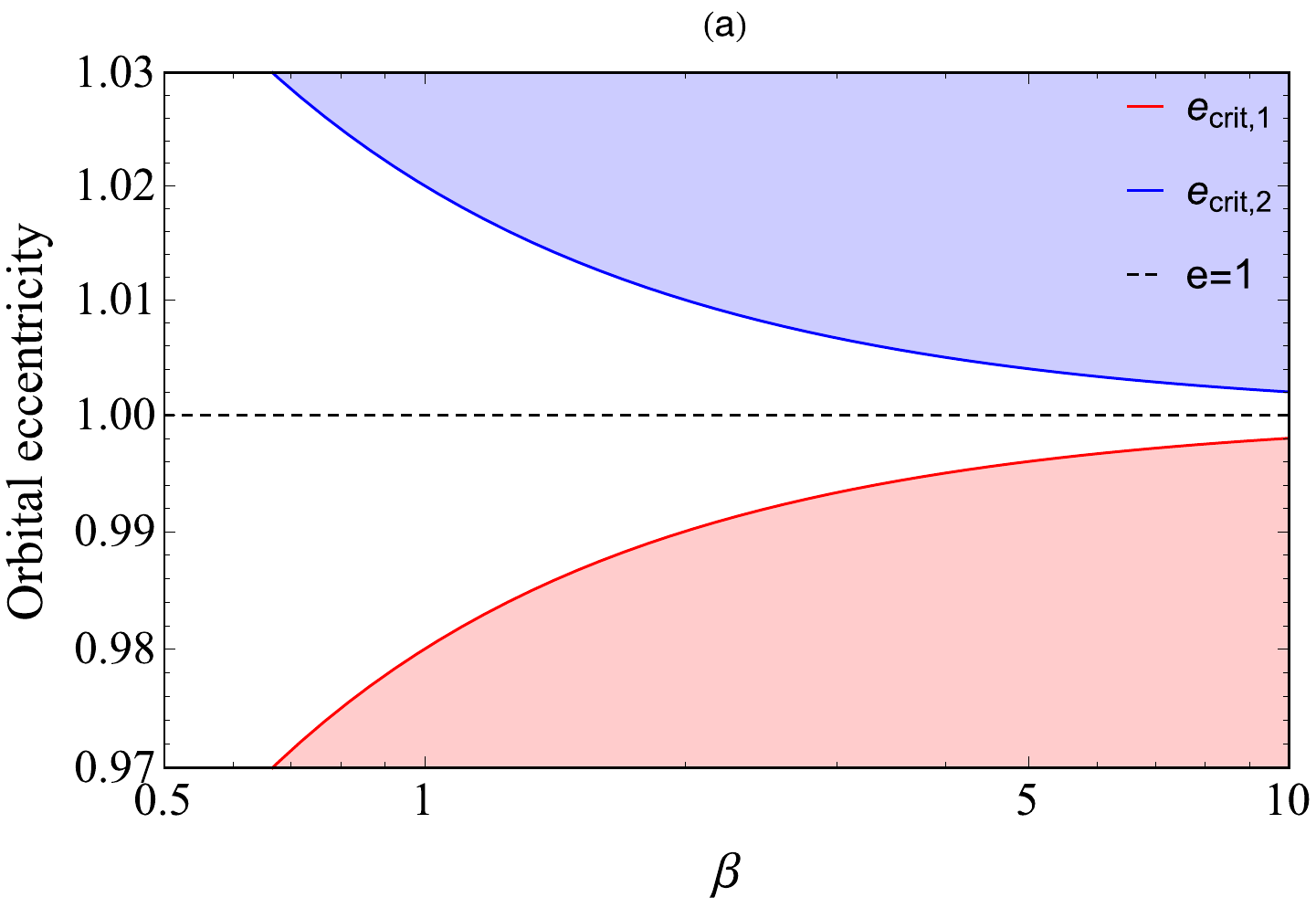}
\includegraphics{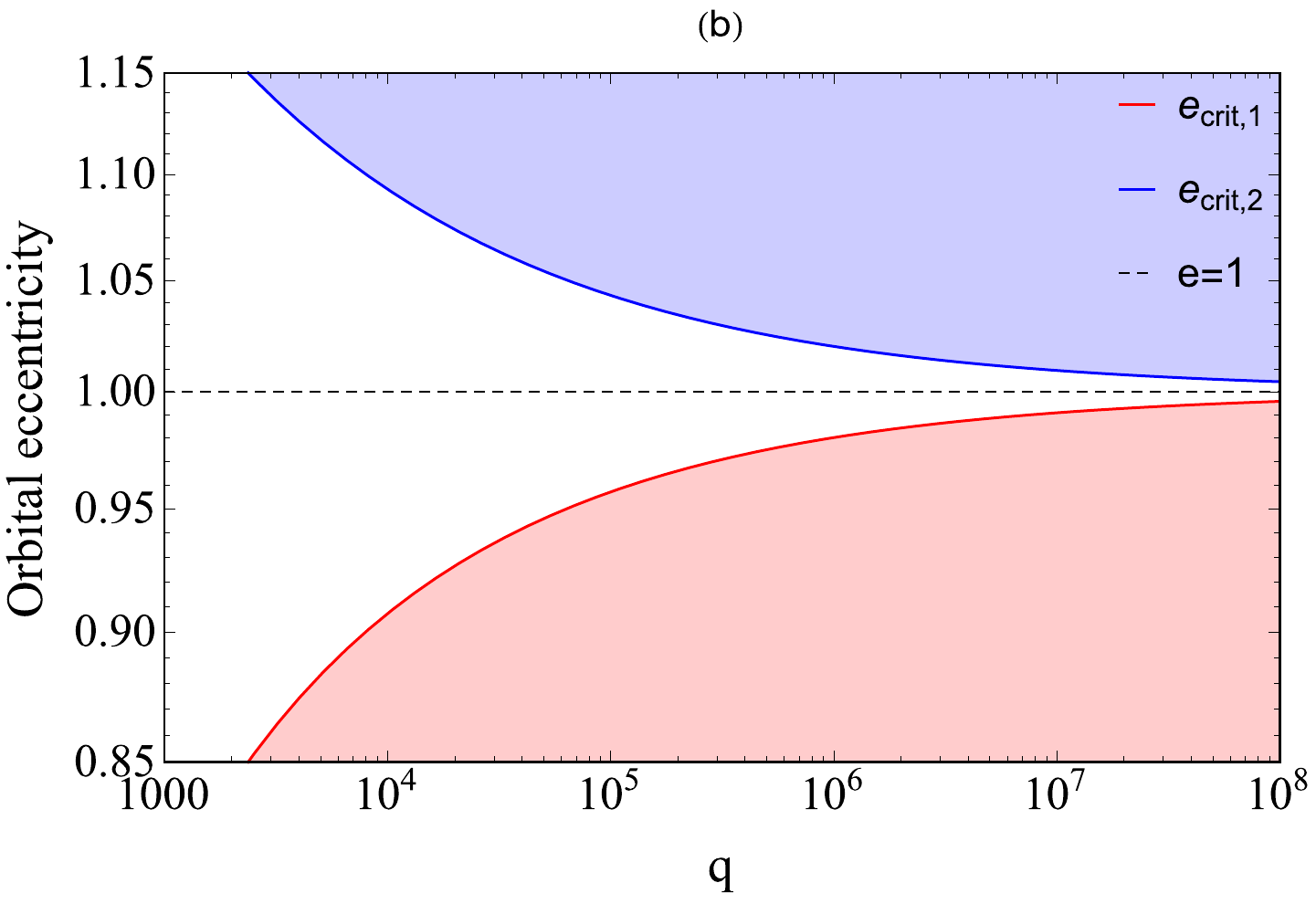}
}
\caption{
Dependence of critical eccentricities on the penetration factor $\beta$ and the 
ratio of the black hole to stellar mass $q=M_{\rm bh}/m_{*}$. 
Panel (a) shows the dependence of the critical eccentricity 
on $\beta$ with the fixed value of $M_{\rm bh}=10^6\,M_\odot$. Panel (b) shows 
the dependence of critical eccentricity on $q$ with the fixed value of $\beta=1$. 
In both panels, red and blue shaded area show the regions of eccentric and hyperbolic 
TDEs, respectively. The white shaded area between the blue and red solid lines show 
the region of parabolic TDEs.
}
\label{fig:class}
\end{figure}

Panel (a) of Figure~{\ref{fig:class}} shows the dependence of critical eccentricities 
on the penetration factor $\beta$ with the fixed value of $M_{\rm bh}=10^6,\,M_\odot$, 
whereas Panel (b) shows the dependence of critical eccentricities on the mass ratio $q$ 
with the fixed value of $\beta=1$. In both panels, red and blue shaded regions show the 
eccentric and hyperbolic TDEs, respectively. The white shaded region between the blue 
and red solid lines show the parabolic TDEs. From the both panels, the higher value of 
$\beta$ and the more massive black holes tend to produce the different type of TDEs by 
the slighter difference of the critical eccentricity. 

%
%

%
%

%
\clearpage
%
\subsection{Modification of mass fallback rates}
%
Following \cite{ek89}, the mass fallback rate is given by
\begin{eqnarray}
\frac{dM}{dt}=\frac{dM}{d\epsilon_{\rm }}\frac{d\epsilon_{\rm }}{dt},
\label{eq:dmdt}
\end{eqnarray}
where $dM/d\epsilon_{\rm }$ is the differential mass distribution of the stellar 
debris with specific energy $\epsilon$. Because the thermal energy of the stellar debris 
is negligible compared with the binding energy, $\epsilon\approx\epsilon_{\rm d}$, where 
$\epsilon_{\rm d}$ is defined as the specific binding energy of the stellar debris by
\begin{equation}
\epsilon_{\rm d}\equiv-\frac{GM_{\rm bh}}{2a_{\rm d}},
\label{eq:ed}
\end{equation}
and by applying the Kepler's third law to it we obtain that 
\begin{equation}
\frac{d\epsilon_{\rm d}}{dt}=\frac{1}{3}(2\pi{GM}_{\rm bh})^{2/3}t^{-5/3}.
\label{eq:dedt}
\end{equation}
Here, we newly assume that 
\begin{eqnarray}
\frac{dM}{d\epsilon_{\rm }}
\equiv
\frac{\eta_{1}(\alpha,a)}{2}\frac{m_*}{\Delta\epsilon}
\left(\frac{-\epsilon_{\rm d}}{\Delta\epsilon}\right)^{\alpha}\,\,\,(\epsilon _{\rm d}<0),
\label{eq:dmde}
\end{eqnarray}
where $\alpha$ is the power law index and $\eta_{1}(\alpha,a)$ is the normalization coefficient 
obtained by the finite integral: 
$\int_{-\Delta\epsilon+\epsilon_{\rm orb}}^{\epsilon_{\rm orb}}\,(dM/d\epsilon_{\rm }^{})\,d\epsilon_{\rm d}^{'}=m_{*}/2$ 
as
\begin{equation}
\eta_{1}(\alpha,a)\equiv(\alpha+1)\left[\left(1+\frac{a_{\rm c}}{a}\right)^{\alpha+1}-\left(\frac{a_{\rm c}}{a}\right)^{\alpha+1}
\right]^{-1}.
\label{eq:eta1}
\end{equation} 
It is required that $\alpha+1$ is greater than zero because of $0\le\eta_{1}(\alpha,a)<\infty$.
If $\alpha=0$ is adopted, equation (\ref{eq:dmde}) 
is reduced, independently of the semi-major axis of the approaching star, to 
the top-hat distribution around zero specific energy: 
$dM/d\epsilon_{\rm }=m_*/(2\Delta\epsilon)$ proposed by \cite{rees88}.
The non-zero value of $\alpha$ represents the effect of the density profile 
of the star on $dM/d\epsilon_{\rm }$. 
In the limit of $a\rightarrow\infty$, equation~(\ref{eq:dmde}) is 
applicable to estimating $dM/d\epsilon_{\rm }$ of the centrally condensed stars on parabolic orbits \citep{lkp09} or 
the partially disrupted stars on parabolic orbits \citep{2013ApJ...767...25G}. 
In the case of eccentric TDEs, $dM/d\epsilon_{\rm }$ has a different distribution 
from the top-hat one (\citealt{hsl13}). This implies that $dM/d\epsilon_{\rm }$ 
of non-parabolic TDEs can deviate from the standard, top-hat distribution. 
Since $dM/d\epsilon_{\rm }$ is a decreasing function of $-\epsilon_{\rm d}$, 
$\alpha$ should be less than or equal to zero. 
The possible range of $\alpha$ is therefore given by $-1<\alpha\le0$.

The specific binding energy of the most tightly bound debris is given by
\begin{eqnarray}
\epsilon_{\rm mtb}=-\Delta\epsilon\pm\frac{GM_{\rm bh}}{2a}, \nonumber 
\end{eqnarray}
where negative and positive signs of the second term of the right-hand side 
originate from the originally approaching stars on eccentric and hyperbolic orbits, 
respectively. It is easily confirmed that $\epsilon_{\rm mtb}$ 
is reduced to be that of precisely parabolic orbit $(e=1)$ in the limit of $a\rightarrow\infty$. 
The orbital period of the most tightly bound debris is proportional to 
$\epsilon_{\rm mtb}^{-3/2}$ from the Kepler's third law:
\begin{eqnarray}
t_{\rm mtb}=\sqrt{\frac{4\pi^2}{GM_{\rm bh}}}a_{\rm c}^{3/2}\left(1\mp\frac{a_{\rm c}}{a}\right)^{-3/2}\,\,\,(a\ge{a}_{\rm c}).
\label{eq:tmtb}
\end{eqnarray}
Substituting equations (\ref{eq:dedt}) and (\ref{eq:dmde}) into equation (\ref{eq:dmdt}) 
with equations~(\ref{eq:ed}) and (\ref{eq:tmtb}), we obtain the modified fallback rate:
\begin{eqnarray}
\frac{dM}{dt}=\frac{\eta_{2}(\alpha,a)}{3}
\left(\frac{m_{*}}{t_{\rm mtb}}\right)
\left(\frac{t}{t_{\rm mtb}}\right)^{-2\alpha/3-5/3},
\label{eq:mdot}
\end{eqnarray}
where $\eta_{2}(\alpha,a)$ is the proportional coefficient defined by 
\begin{equation}
\eta_{2}(\alpha,a)\equiv
(\alpha+1)
\left(
1\mp\frac{a_{\rm c}}{a}
\right)^{\alpha+1}
\left[
\left(1+\frac{a_{\rm c}}{a}\right)^{\alpha+1}
-
\left(\frac{a_{\rm c}}{a}\right)^{\alpha+1}
\right]^{-1}
\label{eq:eta2}
\end{equation}
with the upper and lower signs corresponding to the hyperbolic and eccentric 
orbit cases, respectively. Note that $\eta_{2}(\alpha,a)$ should be greater than or equal to zero 
in order for $dM/dt\ge0$. The relation between $\eta_{2}$ 
and $\eta_{1}$ is given by $\eta_{2}(\alpha,a)=\eta_{1}(\alpha,a)
(1\mp{a_{\rm c}/a})^{\alpha+1}$ from equations (\ref{eq:eta1}) and (\ref{eq:eta2}). 
The possible range of $\alpha$ for a given value of $a$ is therefore 
$-1<\alpha\le0$ in equation (\ref{eq:mdot}).
For both the eccentric and hyperbolic orbit cases, 
the possible range of $a$ is $a_{\rm c}\le{a}<\infty$. 
In the limit of the parabolic orbit ($a\rightarrow\infty$), 
$\eta_{2}(\alpha,a)$ is reduced to $\alpha+1$. 
For the hyperbolic orbit case, $\eta_{2}(\alpha,a)$ is always smaller than unity, and 
$\eta_{2}(\alpha,a_{\rm c})=0$ at the equality of $a={a}_{\rm c}$. 
This equality means that the star approaches the black hole on such a hyperbolic orbit that 
no debris fallbacks after the tidal disruption. 
These arguments imply that the parabolic TDEs are divided into three subclasses: 
marginally eccentric $(e_{\rm crit,1}\le{e}<1$), standard, precisely parabolic $(e=1)$, 
and marginally hyperbolic $(1<e\le{e_{\rm crit,2}})$ TDEs. For useful purpose, we 
summarize the classification of TDEs in Table~1.
%

For marginally eccentric TDEs, the mass fallback rate takes a maximum at $a={a_{\rm c}}$ and $\alpha=0$. 
While the mass fallback rate is 
proportional to $t^{-5/3}$ for $\alpha=0$, it more loosely decays with time for $-1<\alpha<0$. 
For precisely parabolic TDEs ($a\rightarrow\infty$), equation~(\ref{eq:mdot}) reduces to 
$dM/dt=(\alpha+1)/3(m_*/t_{\rm mtb})(t/t_{\rm mtb})^{-2\alpha/3-5/3}$. 
If $\alpha=0$ is adopted, it corresponds to equation (3) of \cite{ek89}: 
$dM/dt=(1/3)(m_*/t_{\rm mtb})(t/t_{\rm mtb})^{-5/3}$. 
For marginally hyperbolic TDEs, the mass fallback rate takes a maximum 
at $a\rightarrow\infty$ and $\alpha=0$, and is close to zero as $a\rightarrow{a_{\rm c}}$. 
It more loosely decays with time for $-1<\alpha<0$ as is the case with the marginally 
eccentric TDEs. For the hyperbolic TDEs ($e_{\rm crit,2}<{e}$), the mass fallback rate should be zero. 
In other words, all the debris mass are unbound to the black hole. The hyperbolic TDEs cannot thus contribute 
to the event rate of the tidal disruption, even if they might occur. The current formula of $dM/dt$ 
significantly underestimates the mass fallback rate of eccentric TDEs ($0\le{e}<e_{\rm crit,1}$), 
because $dM/d\epsilon$ would not follow a simple power law but have more like a Gaussian distribution 
if the specific binding energy of most loosely bound orbit is negative enough beyond 
$-\Delta\epsilon$ \citep{hsl13}. Our conjecture given by equation (\ref{eq:dmde}) should 
be therefore inapplicable to eccentric TDEs.

\begin{deluxetable}{|c|c|c|c|c|}
\rotate
\tablecaption{
Summary for classification of TDEs. The first column represents the type of the TDEs.
The second column shows the condition to classify the TDEs from the viewpoints 
of the orbital eccentricity, $e$, with two critical eccentricities $e_{\rm crit, 1}$ and 
$e_{\rm crit, 2}$ (see equations (\ref{eq:ec1})-(\ref{eq:eclass})). The third column 
shows the condition from the viewpoint of the semi-major axis, $a$, with the critical 
semi-major axis $a_{\rm c}$ (see equations~(\ref{eq:aclass}) and (\ref{eq:ac})). The fourth column denotes 
the possible range of $\alpha$ for each TDE. The 
final column represents the mass fallback rate for each TDE, 
where $\eta_{2}(\alpha,a)$ is the proportional coefficient given by equation (\ref{eq:eta2}) 
and $t_{\rm mtb}$ is the orbital period of the most tightly bound orbit given by equation~(\ref{eq:tmtb}).
}
\tablewidth{0pt}
\tablehead{
Type of TDEs & $e$ & $a$ & $\alpha$ & $dM/dt$
}
\startdata
Eccentric & $0 \le e < e_{\rm crit,1} $ & $0<a<a_{\rm c}$ & $-$ & Intense accretion$^{*}$ \\
\hline 
Marginally eccentric & $e_{\rm crit,1}\le{e}<1$ & ${a}_{\rm c}\le{a}<\infty$ & $-1<\alpha\le0$ & 
$(\eta_{2}(\alpha,a)/3)(m_*/t_{\rm mtb})(t/t_{\rm mtb})^{-2\alpha/3-5/3}$ \\ 
\hline                        
Precisely parabolic & $e=1$ & $a\rightarrow\infty$  & $-1<\alpha\le0$ & $(1/3)(\alpha+1)(m_*/t_{\rm mtb})(t/t_{\rm mtb})^{-2\alpha/3-5/3}$$^{**}$ \\ 
\hline
Marginally hyperbolic & $1<{e}\le{e_{\rm crit,2}} $ & $a_{\rm c}\le{a}<\infty$ 
&  $-1<\alpha\le0$ & $(\eta_{2}(\alpha,a)/3)(m_*/t_{\rm mtb})(t/t_{\rm mtb})^{-2\alpha/3-5/3}$ \\
\hline
Hyperbolic & $e_{\rm crit,2}<{e}$  & $0<a<{a}_{{\rm c}}$ & $-$
& No accretion      \\
\enddata
\tablenotetext{*}{see \cite{hsl13}}
\tablenotetext{**}{It corresponds to equation (3) of \cite{ek89} if $\alpha=0$ is adopted.}
\label{tbl:1}
\end{deluxetable}

%
\clearpage
%

%
\section{N-body experiments}
\label{sec:3}
%

In this section, we present a scaling study of whether there are five (three plus two) types 
of TDEs from the viewpoint of the stellar orbits by N-body experiments, and estimate the 
fractional number of each event rate. All the simulations are performed by using the massively 
parallel $\phi$-GRAPE code \citep{2007NewA...12..357H}, with high performance up to 1.5 
Tflop/s per GPU on our HPC clusters in Beijing (NAOC/CAS) and Heidelberg (ARI/ZAH) 
\citep{ber11,spur12, ber13a, ber13b}. The code is a direct N-body simulation package, 
with a high order Hermite integration scheme and individual block time steps. A direct 
N-body code evaluates in principle all pairwise forces between the gravitating particles, 
and its computational complexity scales asymptotically with $N^2$; however, it is not to 
be confused with a simple brute force shared time step code, due to the block time steps. 
The present code is well-tested and already used to obtain important results in our earlier 
large scale (up to few million body) simulation \citep{khan+12,zhong+14,khan+16,2017ApJ...834..195L}.

%
%
\begin{table*}
\centering
\rotate
  \caption{
Simulation parameters and results. 
The first column shows each model. 
The second, third, fourth, and fifth columns are 
the total number of N-body particles $N$ 
in units of $\rm{K}=1024$, the {accretion radius $\xi_{\rm acc}$}, 
and the number of accreted particles $N_{\rm acc}$, 
and the simulation run time $t_{\rm end}$, respectively. 
The sixth and seventh columns describe the normalized initial 
and final mass of the black hole, respectively. 
The last five columns represent the fractional number 
of the accreted particles on respective orbits in percent figures ($\%$), where $f_{\rm e}$, 
$f_{\rm me}$, $f_{\rm p}$, $f_{\rm mh}$, and $f_{\rm mh}$ are the fractional 
number of the eccentric, marginally eccentric, precisely parabolic, marginally hyperbolic, 
and hyperbolic TDEs, respectively (see Table 1 about the definition of each TDE).
Each mass, size, and time are normalized by 
{\it H\'enon units} ($G=1$, $M_{\rm c}=1$, $r_{\rm c}=1$, and $E_{\rm c}=-1/4$). 
}
  \begin{tabular}{@{}ccccccccccccc@{}}
  \hline
  \hline
Model & {$N/\rm{K}$} & $\xi_{\rm{acc}}$ & $N_{\rm acc}$ & $t_{\rm end}$ 
& $\mu_{\rm ini}$ & $\mu_{\rm{end}}$ & $f_{\rm e}$ & $f_{\rm me}$ & $f_{\rm p}$ & $f_{\rm mh}$ 
& $f_{\rm h}$\\
\hline
\hline
1 &  $128$ & $10^{-5}$ & $449$ & $1500$ 
& $0.01$ & $0.01$ 
& $0$ &$95.6$&$0.2$&$4.2$&$0$  \\
2 &  $128$ & $10^{-4}$ & $1972$ & $700$ 
& $0.01$ & $0.01$ 
& $0$ &$28.9$&$0$&$71.1$&$0$ \\
3 &  $256$ & $10^{-5}$ & $712$ & $1500$ 
& $0.01$ & $0.01$ & $0$ &$75$&$0$&$25$&$0$\\
4 &  $256$ & $10^{-4}$ & $1035$ & $400$ 
& $0.01$ & $0.01$ & $0$ &$19.1$&$0$&$80.9$&$0$ \\
5 &  $512$ & $10^{-5}$ & $693$ & $1000$ 
& $0.01$ & $0.01$ & $0$ &$54.3$&$0$&$45.7$&$0$  \\
6 &  $128$ & $10^{-5}$ & $1141$  & $900$ 
& $0.05$ & $0.05$ & $0$ &$91.1$&$0$&$8.9$&$0$\\
7  & $128$ & $10^{-4}$ & $2433$ & $500$ 
& $0.05$ & $0.05$ & $0$ & $29.82$ & $0.08$ & $70.1$ & $0$  \\
8 &  $256$ & $10^{-5}$ & $1318$ & $700$ 
 & $0.05$ & $0.05$ & $0$ &$67.5$&$0$&$32.5$&$0$  \\
9 & $256$ & $10^{-4}$ & $3763$  & $500$ 
& $0.05$ & $0.05$ & $0$ &$22.3$&$0.0$&$77.7$&$0$ \\ 
10 &  $512$ & $10^{-5}$ & $1854$ & $600$ 
& $0.05$ & $0.05$ & $0$ &$45.95$&$0.05$&$54$& $0$\\
11 &  $128$ & $10^{-5}$ & $1171$  & $2200$ 
& $0.01$ & $0.019$ & $0$ &$54$&$0$&$46$&$0$   \\
12 & $128$ & $10^{-4}$ & $8288$ & $2450$ 
& $0.01$ & $0.073$ & $0.2$ &$87.8$&$0$&$12$&$0$   \\
13 & $128$ & $10^{-3}$ & $16620$ & $2200$ 
& $0.01$ & $0.14$ & $0.6$ &$22.8$&$0$&$76.4$&$0.2$   \\
14 &  $256$ & $10^{-5}$ & $1627$ & $2522$ 
& $0.01$ & $0.018$ & $0$ &$39.3$&$0$&$60.7$&$0$   \\
15 &  $256$ & $10^{-4}$ & $6651$ & $1400$ 
& $0.01$ & $0.035$ & $0$ &$13.5$&$0$&$86.5$&$0$   \\
\hline
\end{tabular}
\label{tbl:2}
\end{table*}


%
\subsection{Method}
%

We use the same simulation method as \cite{zhong+14} here. 
In our simulations, $G=1$, $M_{\rm c}=1$, $r_{\rm c}=1$, $E_{\rm c}=-1/4$ 
{({\it H\'enon units})} are adopted for useful purpose {\citep{1971Ap&SS..13..284H,hm86}}, 
where $G$, $M_{\rm c}$, $r_{{\rm c}}$, and $E_{\rm c}$ are the gravitational 
constant, the total mass of the cluster, {the virial radius} and energy 
of the star cluster, respectively. 

We choose different values of $N$ and {introduce the normalized accretion 
radius $\xi_{\rm acc}\equiv{r}_{\rm acc}/r_{\rm c}$} to evaluate
 the physical scaling behavior of our system and extrapolate to the real system.
 Here $N = M_{\rm c}/m$ is the particle number, which defines the ratio between particle mass $m$
 and total mass of the system $M_{\rm c}$. Note that $m$ does not have to be identical with the
 stellar mass. Currently adopted $N$ is from 128K to 512K (see Table \ref{tbl:2}), where 
 we define $1K=1024$ due to technical reason through this paper, and  a Plummer model is adopted 
for the initial stellar distribution \citep{aa74}.

 The normalized accretion radius $\xi_{\rm acc}$ is another dimensionless number which
 defines a radius at which simulation particles are to be disrupted by tidal forces of a
 central black hole, relative to the virial radius of our system, which is used as standard unit. 
 Extrapolation to the real system means that $N$ is approaching real particle (star) numbers
 (like say $10^8$ in galactic nuclei) and $\zeta\equiv{r}_{\rm acc}/r_{\rm t}$ is close to $1$ 
 at the same time. We also have a third dimensionless parameter in our models, which is 
 $\mu = M_{\rm bh}/M_{\rm c}$. From the standard relations between galactic bulges and 
 central massive black holes \citep{mtr98,mm13}, it should be up to $\sim0.006$. However, 
 we choose higher values because we only simulate part of the central star cluster mass; 
 $\mu=(0.01, 0.05)$.
 
In our simulations, there are two type of sink particles; one is that the black hole is fixed with 
no accretion. In this case, the stars entering inside a finite accretion radius, corresponding to 
the tidal disruption radius, contribute no growth of the black hole and add no linear momentum to 
the black hole, and are right away removed from the stellar system. It looks unphysical but 
is enough to test which orbit the stars are tidally disrupted on. 
Another type is that the black hole particle simply gains the masses of the removed stars without 
adding their linear momentum. Once the star comes into the tidal disruption radius, it will be 
removed from the stellar system. The similar approaches we already implement in $\phi$-GRAPE/GPU 
code and well tested against all the energy and momentum conservations in our earlier works 
\citep{2012ApJ...758...51J, 2016MNRAS.460..240K}. 
The initial density profile of the Plummer model has a central flat core, which adjusts to 
the gravity of the central back hole during a few dynamical orbits, as is the case of \citet{zhong+14}.
In any case, all the stars have equal mass and forms no binary 
stars through the simulations. We adopt three fixed accretion radii in N-body units: 
$\xi_{\rm acc}=10^{-3}$, $10^{-4}$, and $10^{-5}$. We also run the model with $\xi_{\rm acc}=5\times10^{-5}$, 
which are used to extrapolate our simulation models to a realistic system (see section~\ref{sec:dis}).
The accretion radius we used here are larger than the tidal disruption radius, 
typically boosted by a factor of $10^{3-4}$ for the SMBH cases, 
because of our scaling requirements. We discuss this in Section~\ref{sec:dis}.

Table~{\ref{tbl:2}} shows the simulation parameters and results.
The first column shows each simulated model. The second, third, 
fourth, and fifth columns are the total number of N-body particles $N$ 
in units of $\rm{K}=1024$, the normalized accretion radius $\xi_{\rm acc}$, 
and the number of accreted particles $N_{\rm acc}$ entering inside the 
accretion radius, and the simulation run time $t_{\rm end}$, respectively. 
The sixth and seventh columns describe the initial and 
final mass of the black hole normalized by $M_{\rm c}$, respectively. 
Models 1-5 show the simulations for $\mu=0.01$ case. 
Models 6-10 represent those for $\mu=0.05$. Models 11-15 show the 
simulations for the growing black hole case with the initial value of $\mu=0.01$. 
For the non-growing black hole case (Models 1-10), the simulations has been 
stopped when roughly one (or a few) percent of the stars are accreted, 
which is shorter than the half-mass relaxation time given by equation (4) 
of \citet{zhong+14}. This is because no black hole growth in spite of accretion 
can produce the artificial expansion of the cluster, leading to the unphysical effects. 
For the growing black hole case, $t_{\rm end}$ is limited for each model of Models 11-15 
only by the computational resources, but is longer than the half-mass relaxation time 
except for Model 15. In the last five columns, we show the fractional number 
of the accreted particles on respective orbits in percent figures ($\%$), where $f_{\rm e}$, 
$f_{\rm me}$, $f_{\rm p}$, $f_{\rm mh}$, and $f_{\rm h}$ are the fractional number of 
the eccentric, marginally eccentric, precisely parabolic, marginally hyperbolic, and 
hyperbolic TDEs, respectively (see Table 1 about the definition of each TDE).

%
\subsection{Results}
%

In this section, we describe the results of N-body simulations. 
The rate of accreted stars is defined by 
$\langle\dot{M}_{\rm acc}\rangle=(M_{\rm{c}}/t_{\rm end})(N_{\rm acc}/N)$, 
which is estimated to be $10^{-5}$ to $10^{-6}$ in simulation unit for all the 
models. All of our simulations do not reach the steady state for the rate.
This is because the state of the loss cone, which controls the rate of accreted 
stars, changes with time. In the early phase, the loss cone is full as well as a 
density cusp forms around the central black hole, leading to an enhancement 
of the accretion rate. On the other hand, the empty loss cone leads to the 
reduction of the accretion rate in the late phase when a few half-mass relaxation 
time elapse (see also Figure~1 of \citealt{zhong+14}).

\begin{figure}[!ht]
\resizebox{\hsize}{!}{
\includegraphics[width=8cm]{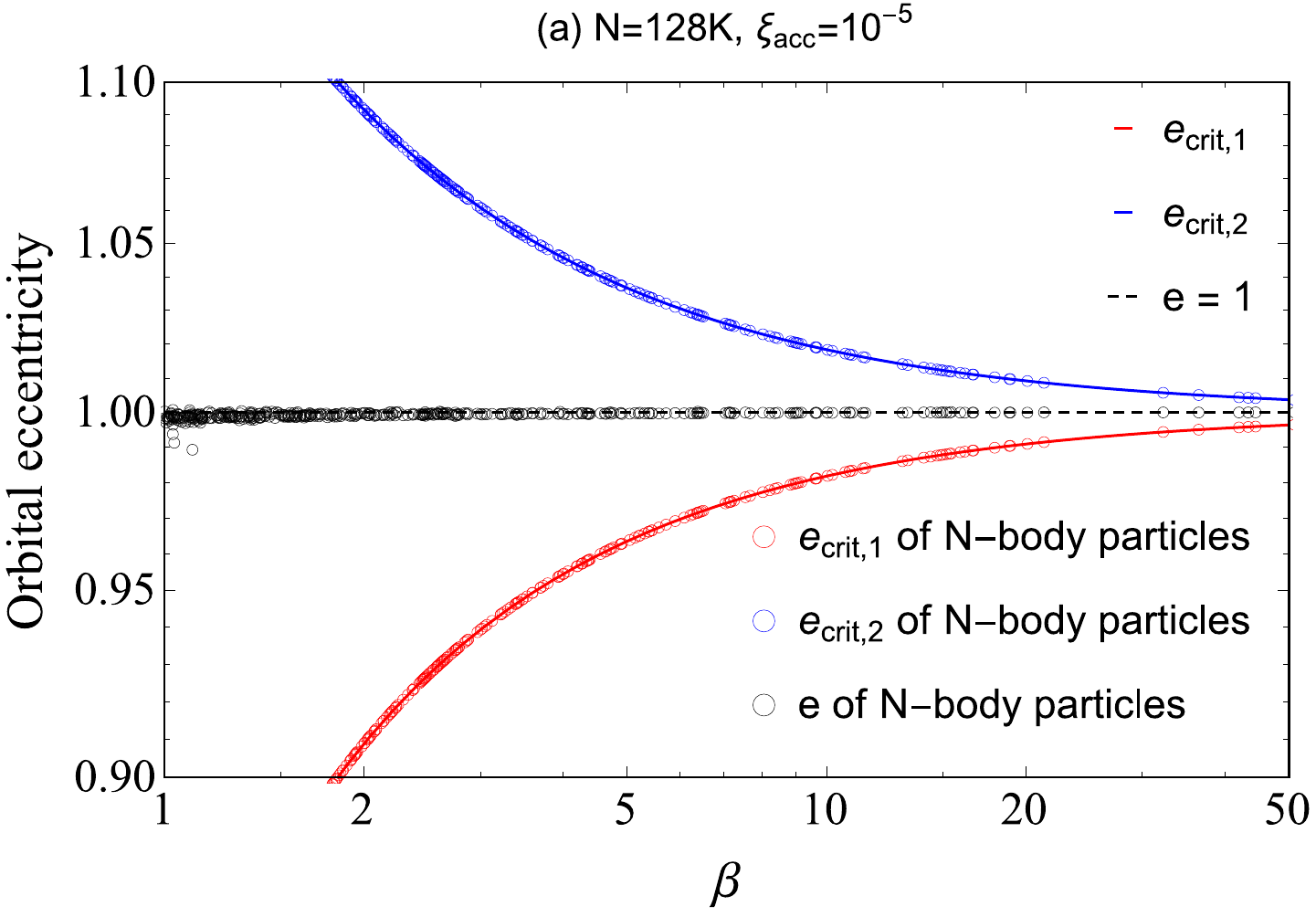}
\includegraphics[width=8cm]{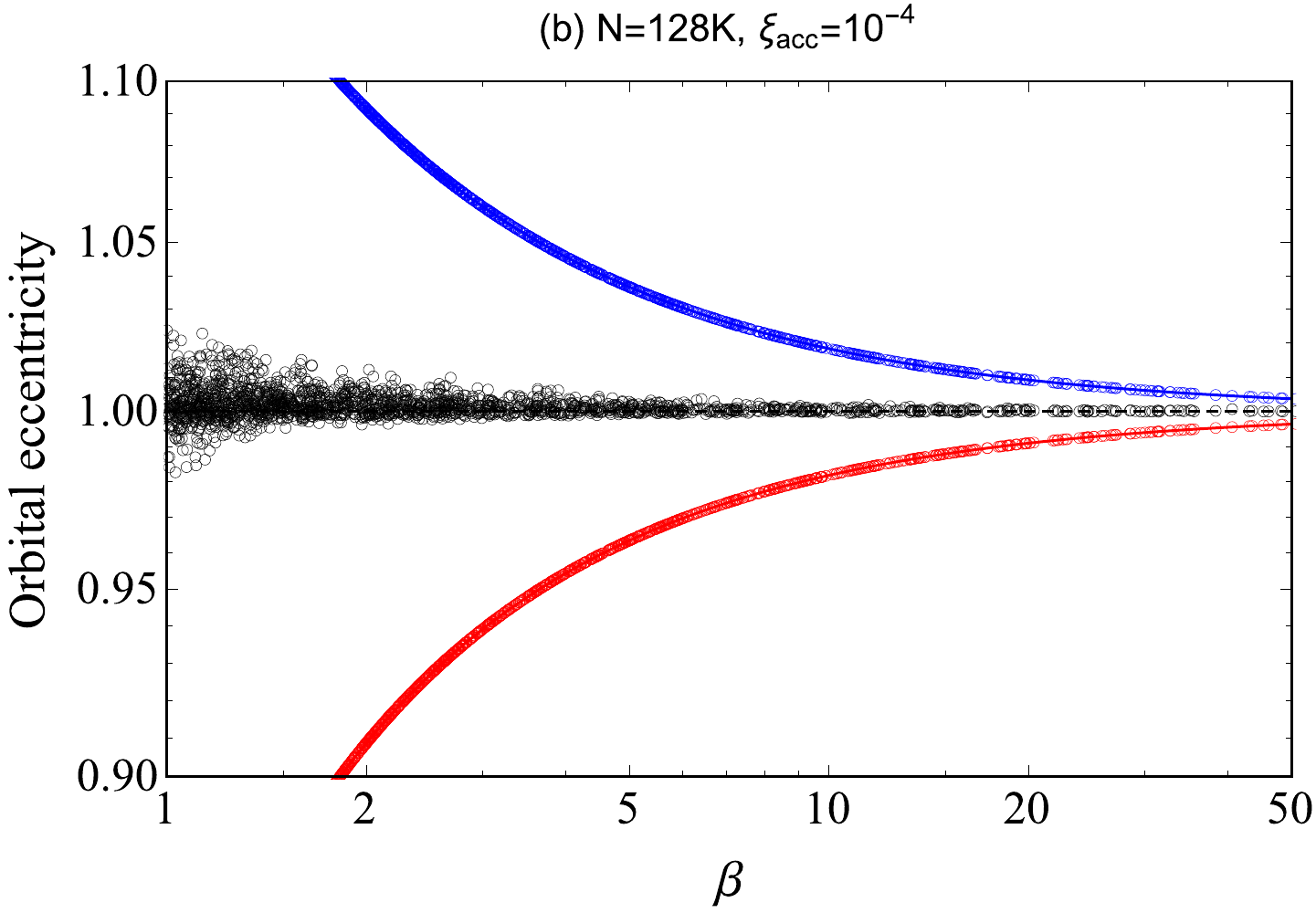}
}\\
\resizebox{\hsize}{!}{
\includegraphics[width=8cm]{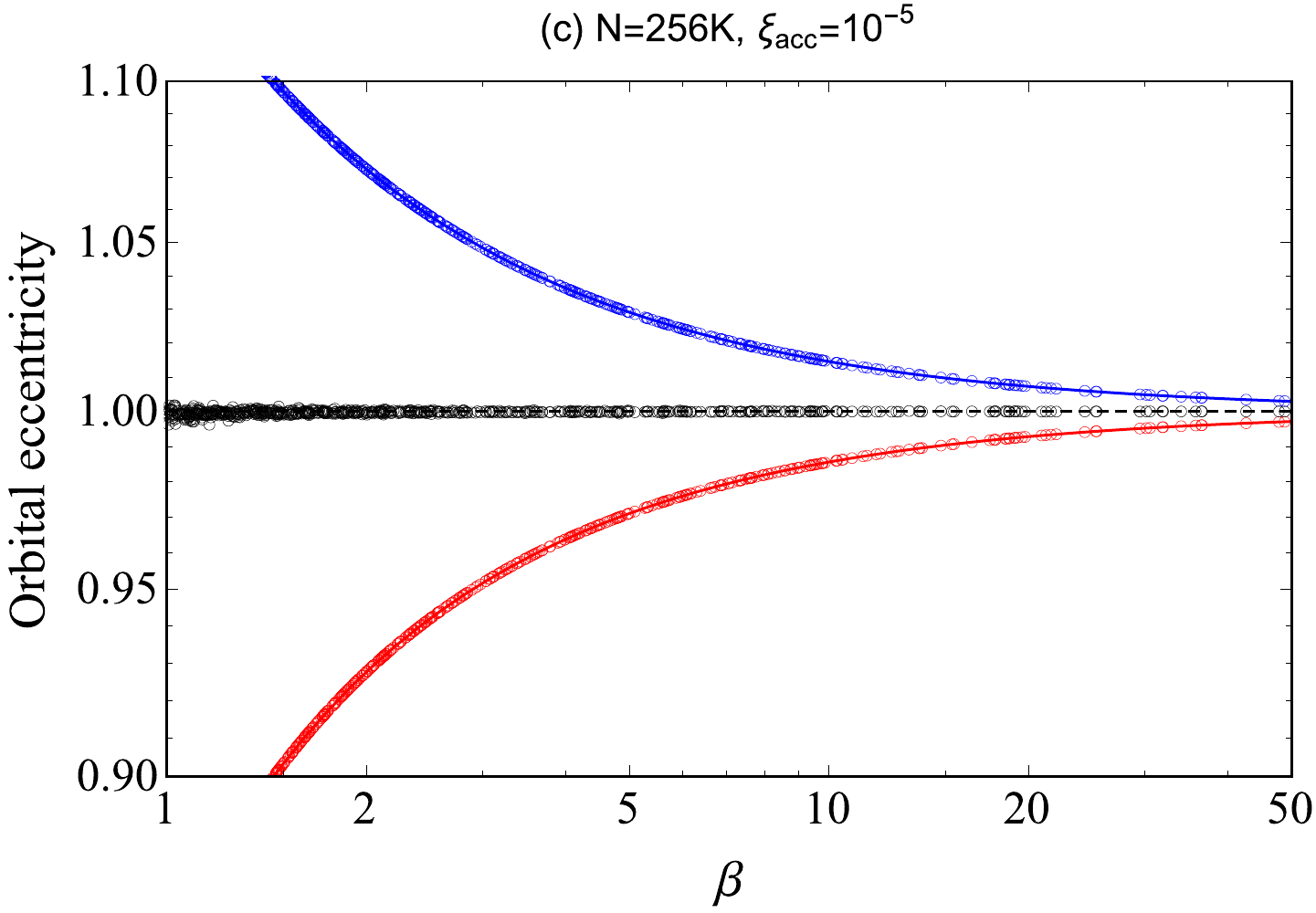}
\includegraphics[width=8cm]{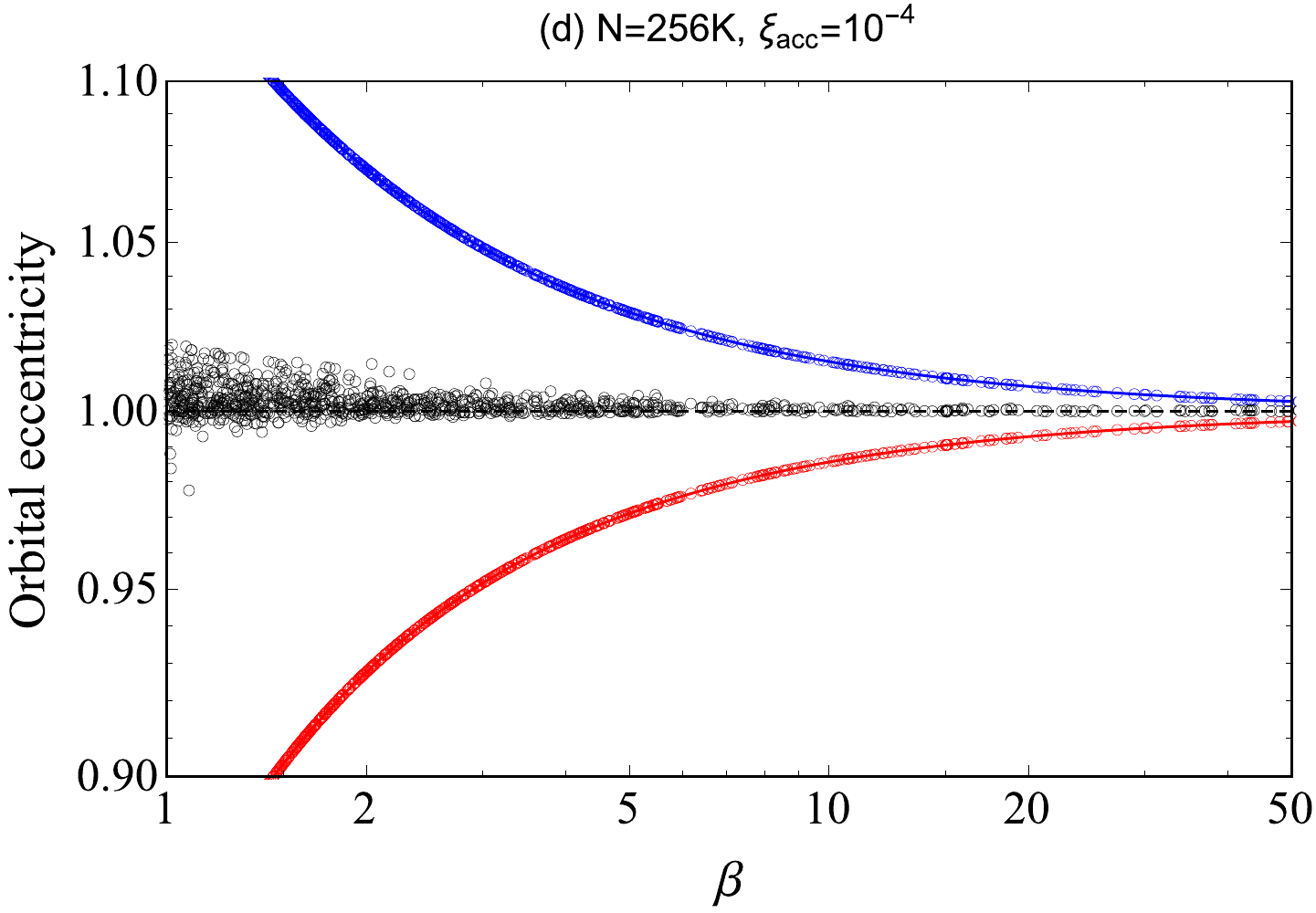}
}\\
\includegraphics[width=8cm]{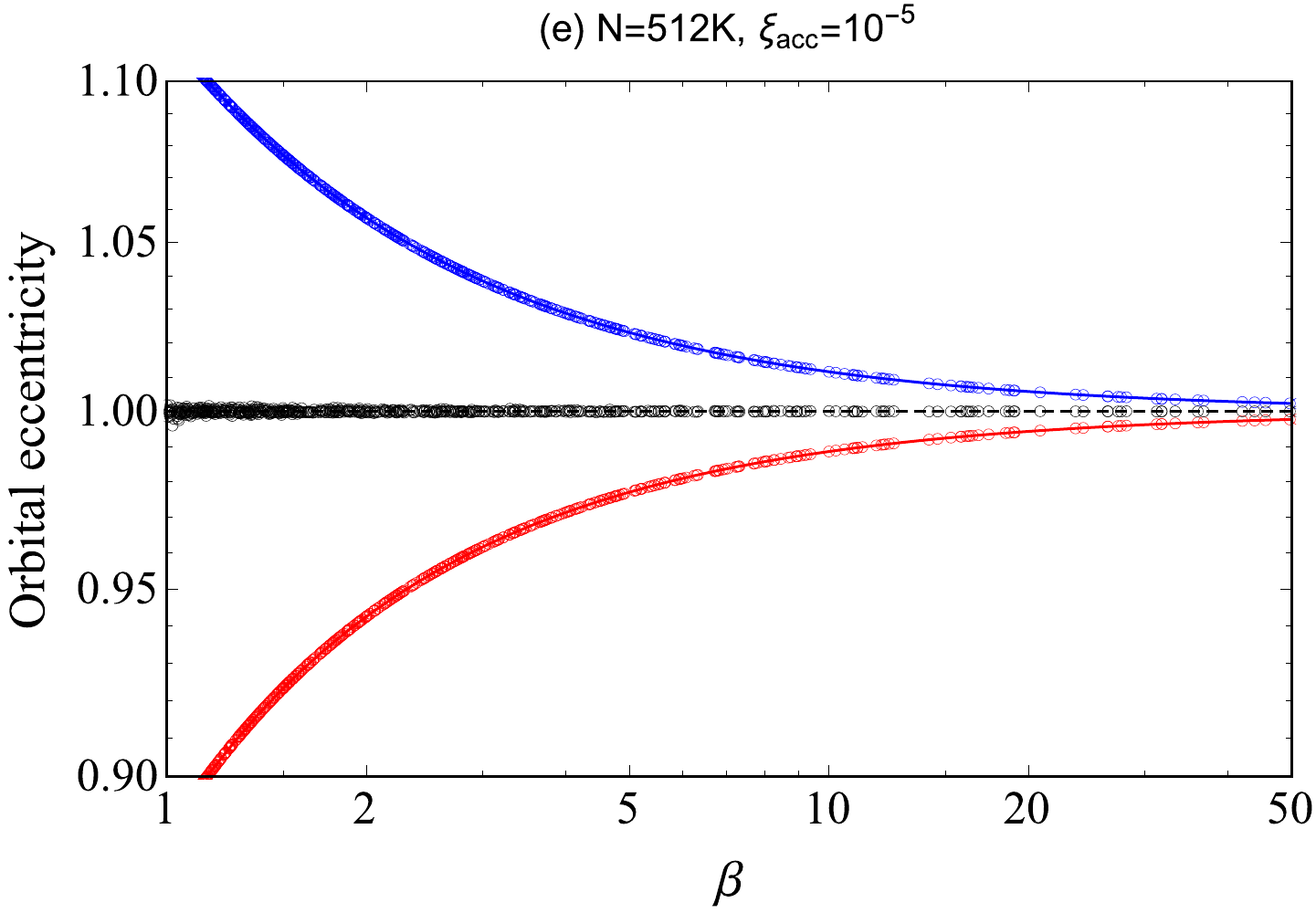}
\caption{
Dependence of critical and orbital eccentricities on the penetration factor 
$\beta$ in the case of $\mu=0.01$ (Models 1-5).
While red and blue small circles represent the critical eccentricities of each 
accreted N-body particle for eccentric and hyperbolic TDEs, respectively. 
The black small circles show the orbital eccentricities of the accreted N-body particles. 
The black dashed line denotes $e=1$, while the red and blue solid lines denote the 
corresponding analytically expected critical eccentricities. 
}
\label{fig:ecrit-sim1}
\end{figure}

\begin{figure}[!ht]
\resizebox{\hsize}{!}{
\includegraphics[width=8cm]{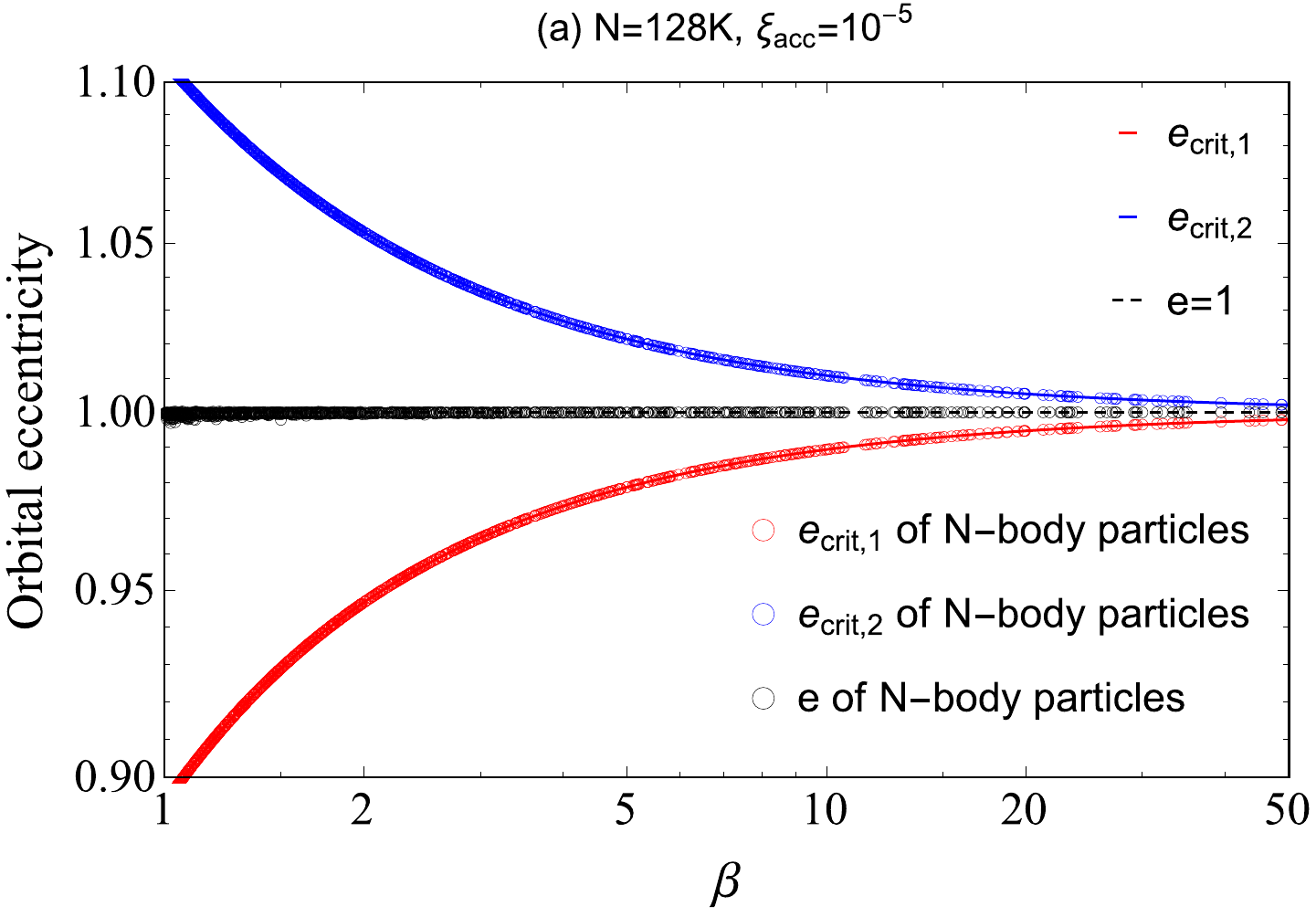}
\includegraphics[width=8cm]{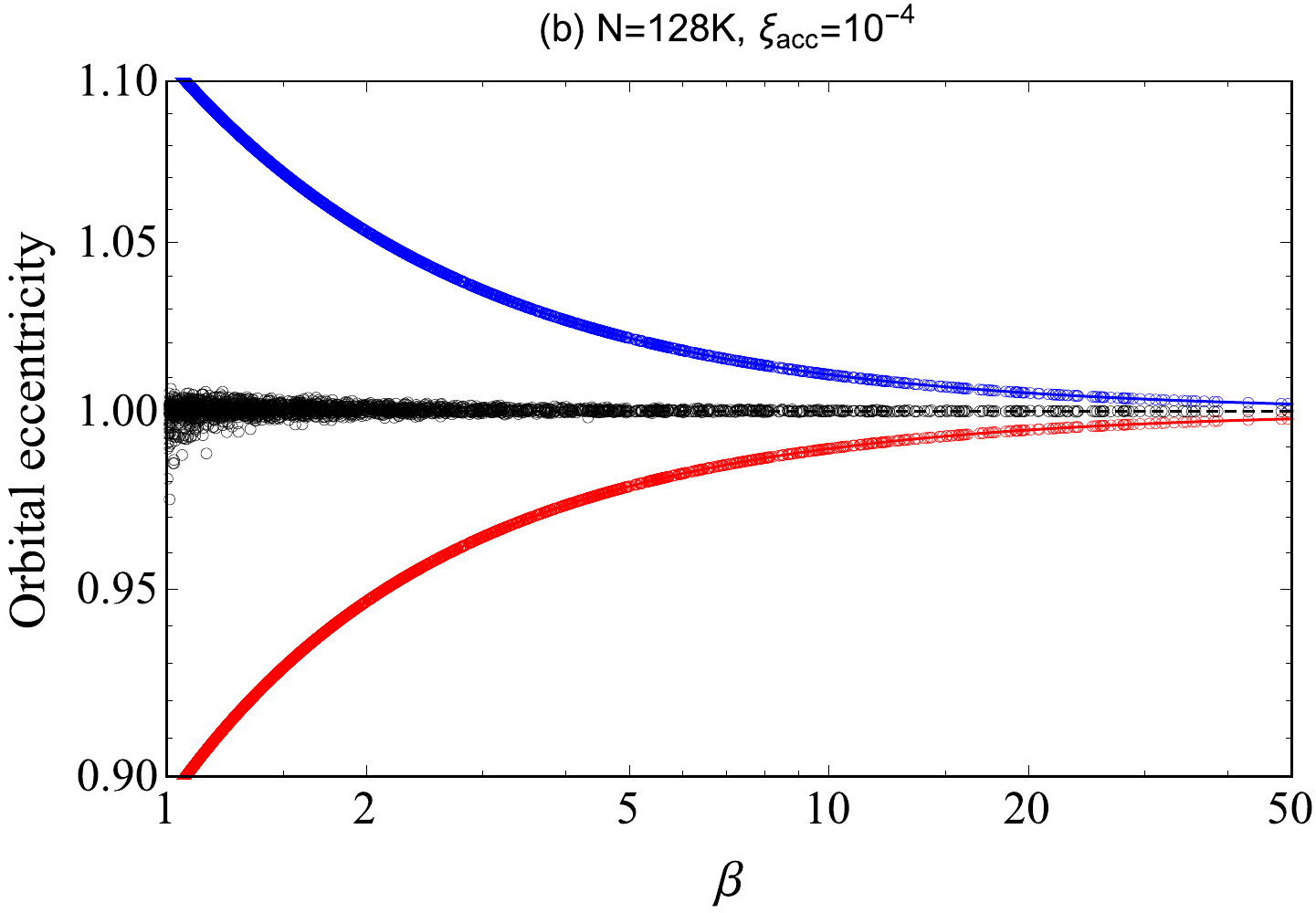}
}\\
\resizebox{\hsize}{!}{
\includegraphics[width=8cm]{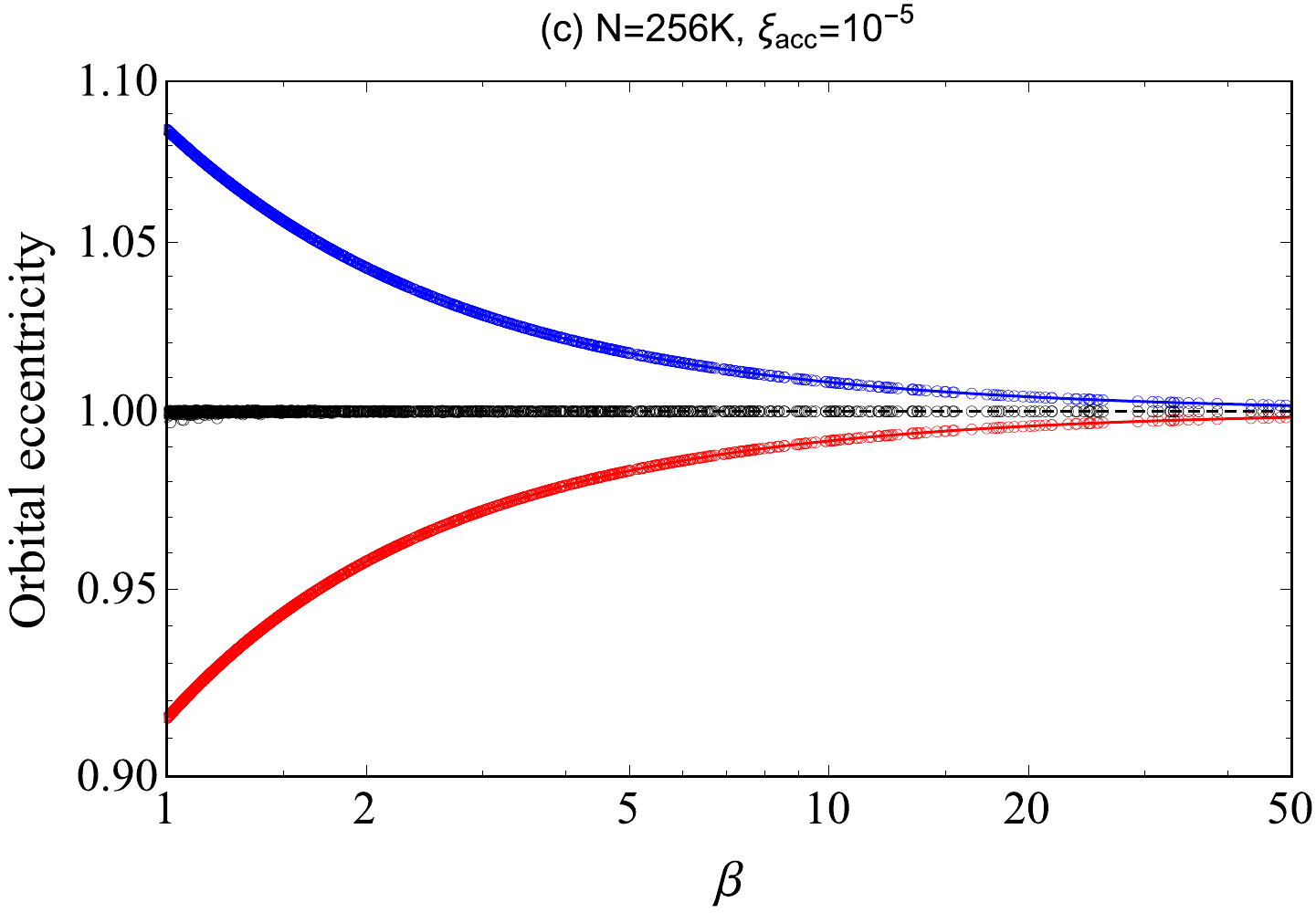}
\includegraphics[width=8cm]{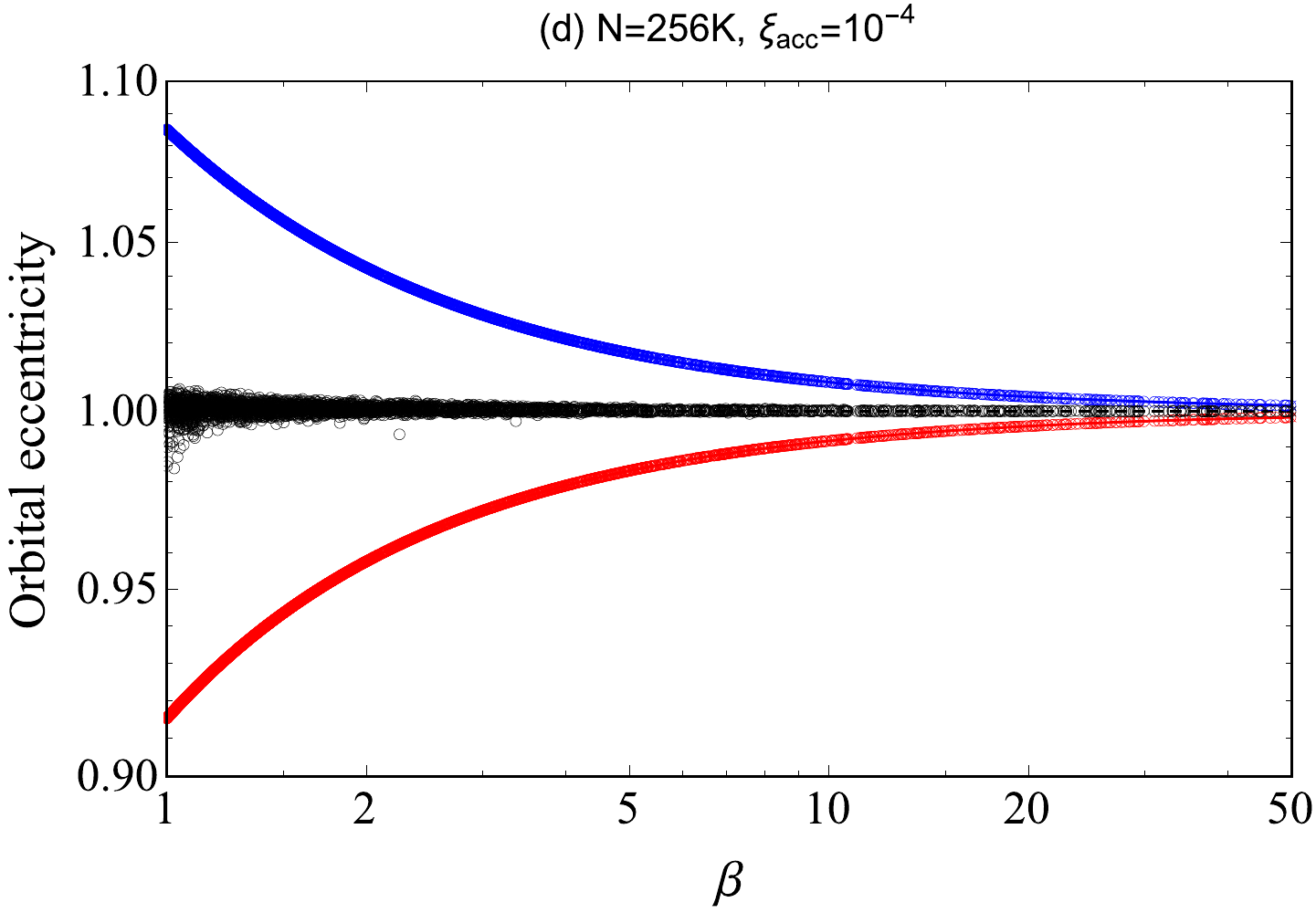}
}\\
\includegraphics[width=8cm]{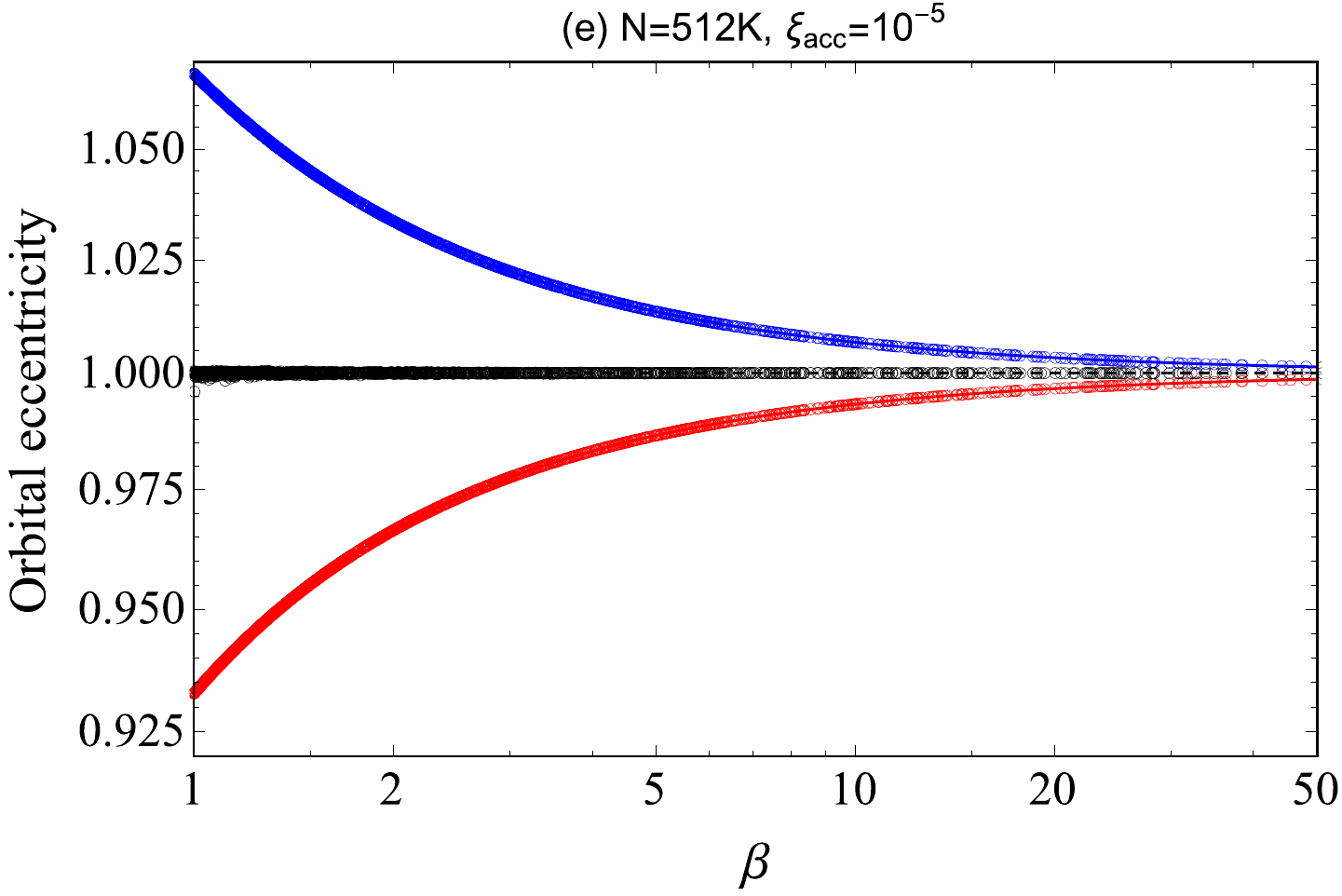}
\caption{
The same formats as Fig.~\ref{fig:ecrit-sim1} but for 
the case of {$\mu=0.05$} (Models 6-10).
}
\label{fig:ecrit-sim2}
\end{figure}

\begin{figure}[!ht]
\resizebox{\hsize}{!}{
\includegraphics[width=8cm]{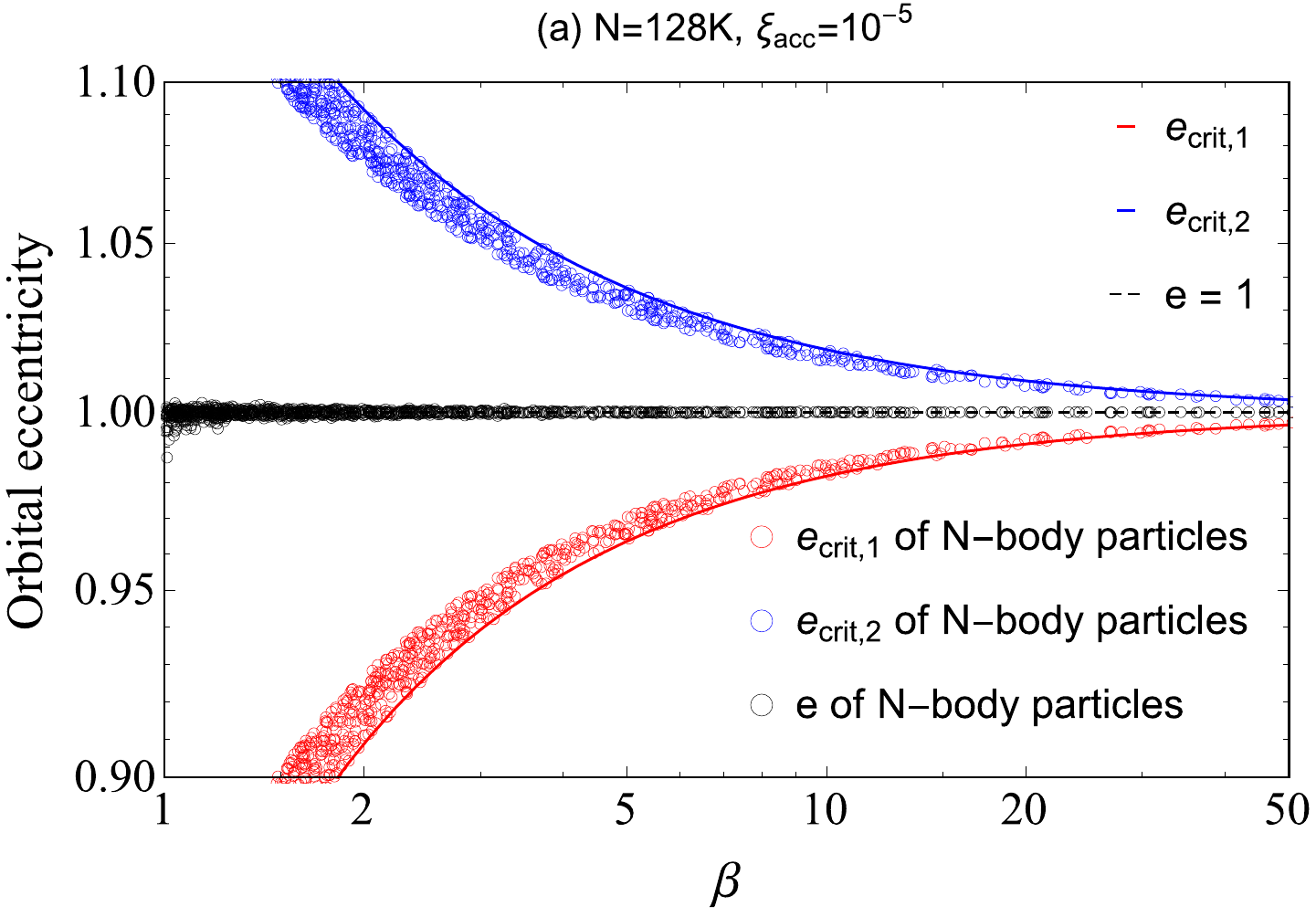}
\includegraphics[width=8cm]{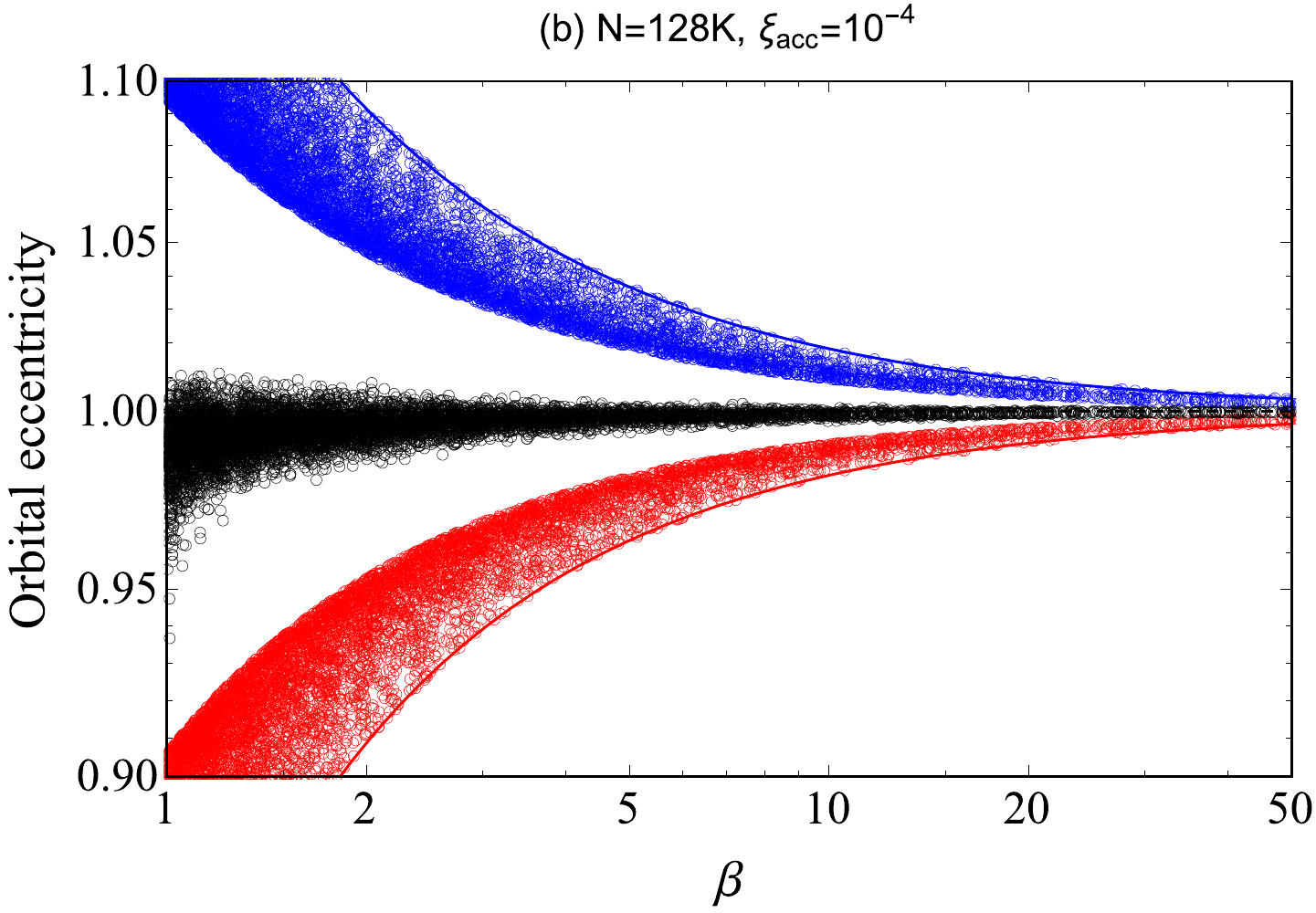}
}\\
\resizebox{\hsize}{!}{
\includegraphics[width=8cm]{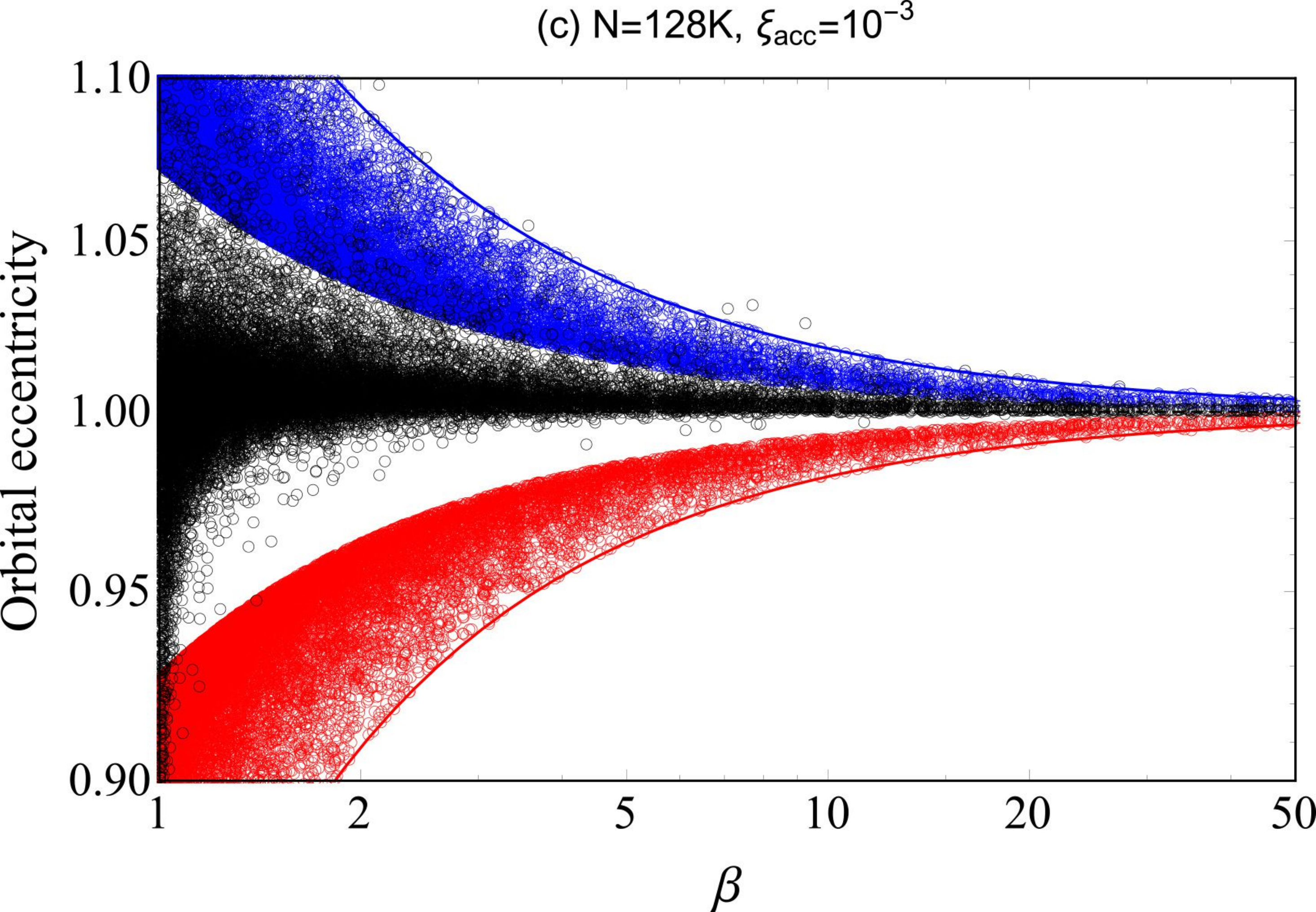}
\includegraphics[width=8cm]{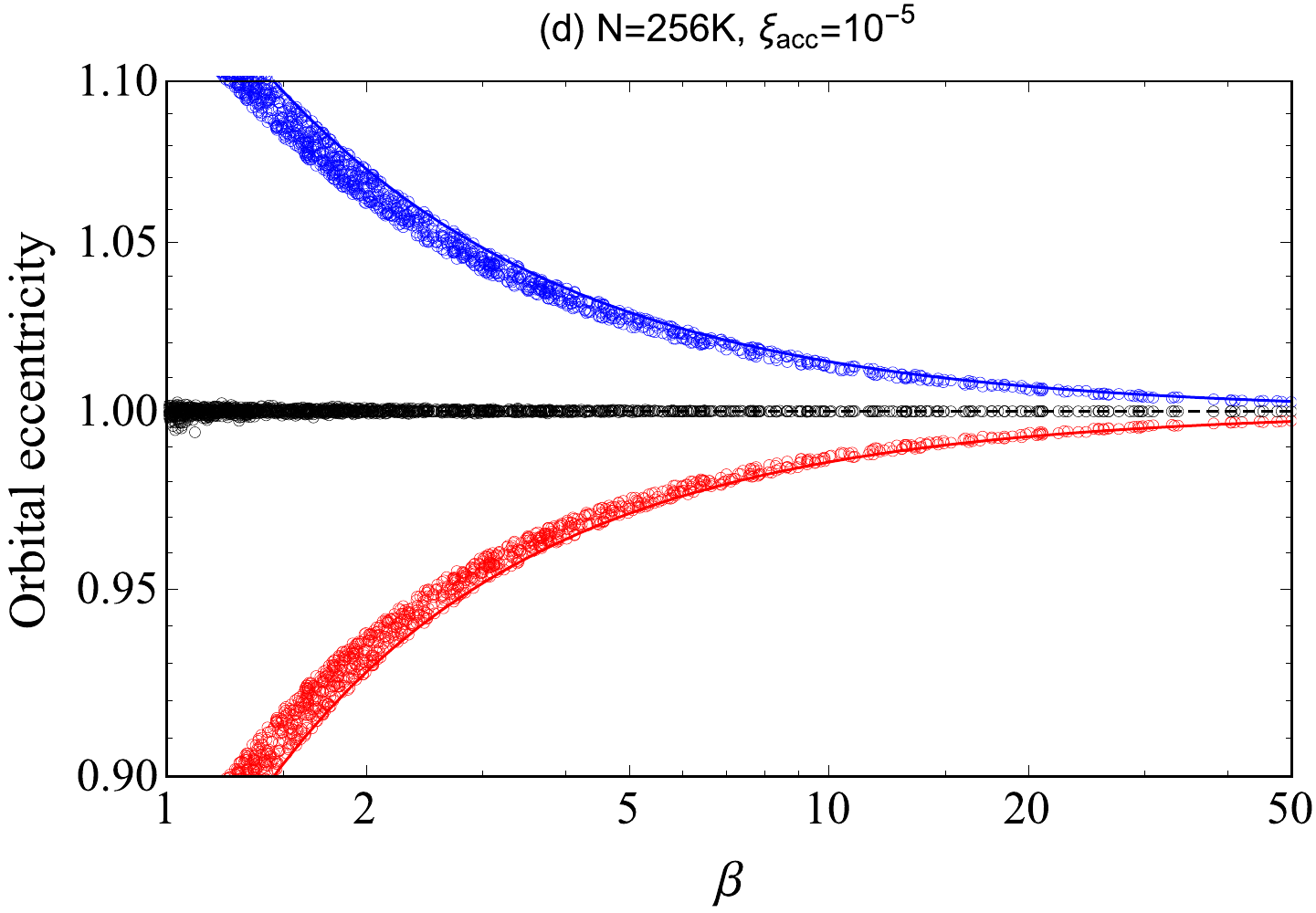}
}\\
\includegraphics[width=8cm]{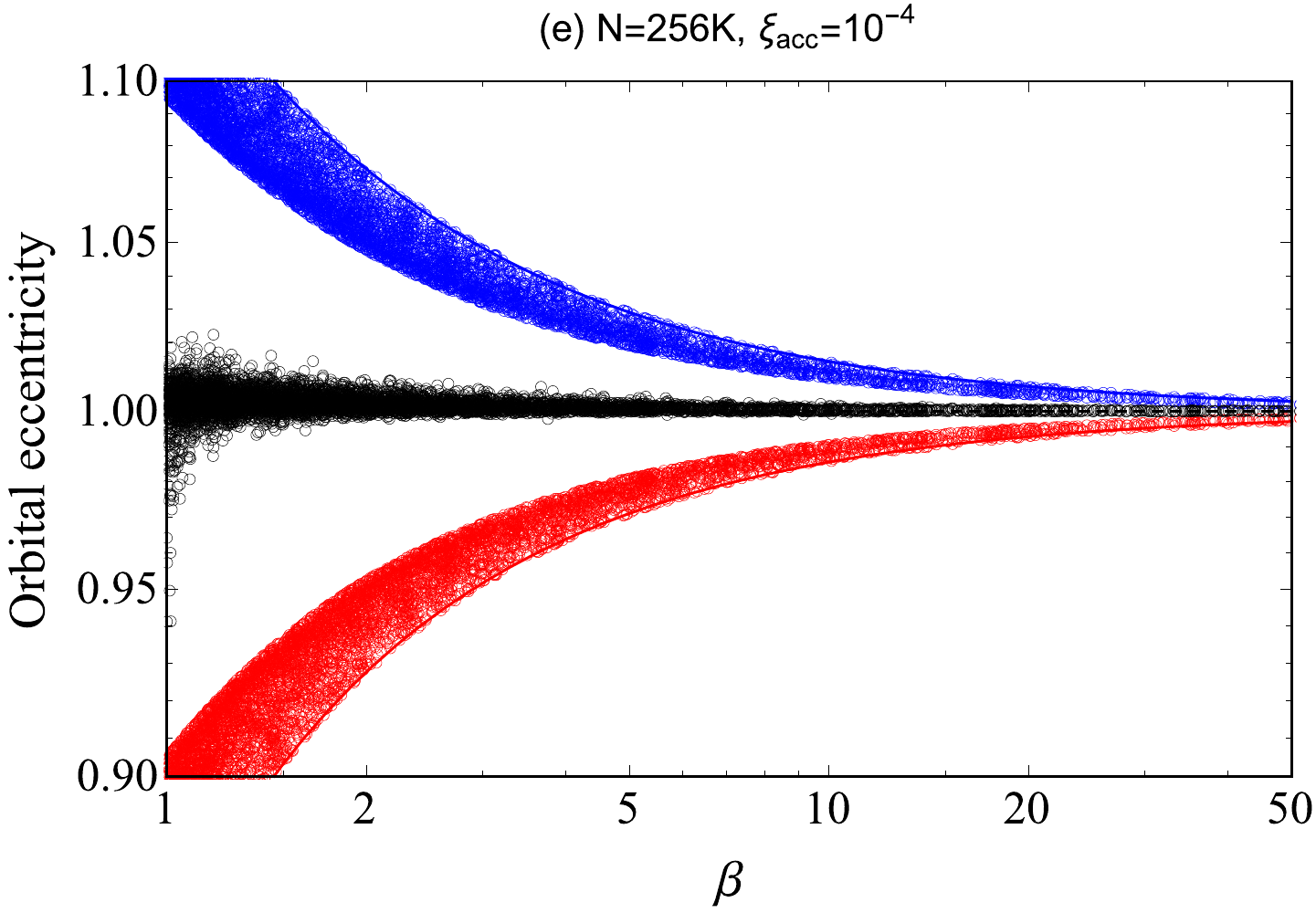}
\caption{
The same formats as Fig.~\ref{fig:ecrit-sim1} but for 
the case of the growing black hole initially from $\mu=0.01$ (Models 11-15).
}
\label{fig:ecrit-sim3}
\end{figure}

Figures~\ref{fig:ecrit-sim1}-\ref{fig:ecrit-sim3} show the dependence 
of the orbital eccentricity of the N-body particles, which accretes inside 
the accretion radius, on the penetration factor, $\beta$. 
Hereafter, we call it $e-\beta$ distribution of the accreted stars. 
In these figures, the black small circles represent $e-\beta$ distribution 
of the accreted N-body particles, whereas the black dashed line denotes $e=1$. 
The red and blue solid lines show the critical eccentricities that are 
analytically expected from equations (\ref{eq:ec1}) and (\ref{eq:ec2}) 
with the fixed value of the mass ratio of the black hole to N-body particles, 
while the red and blue small circles show the two critical eccentricities of each 
N-body particle, which are numerically determined by substituting both 
$\beta$ of each N-body particle and the mass ratio of the black hole to 
N-body particles into equations (\ref{eq:ec1}) and (\ref{eq:ec2}).
%

%
Figure~\ref{fig:ecrit-sim1} shows the $e-\beta$ distribution in Models 1-5.
We confirm that the numerically calculated critical eccentricities are in good 
agreement with the analytically expected ones. In addition, almost all the 
accreted particles are distributed closely around $e=1$ between two critical 
eccentricities. This means that eccentric and hyperbolic TDEs extremely rarely occur. 
For $\xi_{\rm acc}={10^{-4}}$ cases (Models 2 and 4), the N-body particles are clearly 
distributed in the range of $e_{\rm crit,1}<e<e_{\rm crit,2}$. 
We note that a significant faction of the accreted particles will undertake the 
marginally eccentric and marginally hyperbolic TDEs. We also note from 
Figure~\ref{fig:ecrit-sim2} that the $e-\beta$ distributions of Models 6-10 
qualitatively correspond to those of Models 1-5.

The $e-\beta$ distribution of the accreted stars in the growing black hole case 
is different from that of the fixed black hole mass case mainly in following two points. 
Figure~\ref{fig:ecrit-sim3} represents the $e-\beta$ distribution of the accreted 
stars in Models 11-15. The first point is that the numerically calculated critical 
eccentricities deviate from the analytically expected ones. In Models 11-15, the 
black hole mass increases with time by the accreted particles during the simulations. 
As seen in panel (b) of Figure~\ref{fig:class}, both of two critical eccentricities is closer 
to unity (e=1) with the growth of the black hole particle. 
The second point is that the number of more strongly bound 
N-body particles is larger than the fixed black hole case by comparison between 
panel (d) of Figure~\ref{fig:ecrit-sim1} and panel (d) of Figure~\ref{fig:ecrit-sim3}. 
This is because the deeper gravitational potential of the black hole capture more 
the particles at a same distance from the black hole as that of the non-growth case. 
Moreover, some N-body particles clearly have the orbital eccentricity beyond 
the two critical eccentricities, as seen in panel (e). This is because the largest cross 
section makes it possible for the particles with the larger angular momentum to 
accrete onto the black hole than {$\xi_{\rm acc}=10^{-4}$ and $\xi_{\rm acc}=10^{-5}$} 
cases.

Finally, let us see how the fraction of accreted particles is assigned to 
the types of eccentric, marginally eccentric, precisely parabolic, marginally 
hyperbolic, and hyperbolic TDEs. 
Because $f_{\rm e}$, $f_{\rm p}$, and $f_{\rm h}$ are very tiny 
as seen in the last five columns of Table~\ref{tbl:2}, the eccentric, 
precisely parabolic, and hyperbolic TDEs are extremely rare events. 
Almost all of the accreted particles originate from 
N-body particles on marginally eccentric ($e_{\rm crit,1}\le{e}<1$) 
or marginally hyperbolic orbits ($1<e\le{e_{\rm crit,2}}$).
We also find from the last five columns of Table~\ref{tbl:2} that 
the ratio of $f_{\rm me}$ to $f_{\rm mh}$ drastically changes. 
This can be interpreted as follows: 
while all of the stars inside the influence radius of the central black hole, 
$r_{\rm h}=GM_{\rm bh}/\sigma^{2}$, where $\sigma$ is the cluster's 
velocity dispersion, are bounded to the black hole, the stars outside the 
influence radius are unbound to the black hole. According to the loss cone 
theory (\citealt{fr76}; see also \citealt{2013CQGra..30x4005M} for a review), 
the stars are supplied to the black hole mainly from the critical radius, $r_{\rm crit}$, 
where the opening angle of the loss cone angle $\theta_{\rm lc}\approx\sqrt{r_{\rm t}/r_{\rm crit}}$ 
for $r\lesssim{r_{\rm h}}$ is equal to the diffusion angle $\theta_{\rm D}\propto\sqrt{\ln{N}/N}$. 
Because $r_{\rm crit}$ is proportional to $(N/\ln{N})r_{\rm t}$, it depends on each model. 
If $r_{\rm crit}$ is smaller than $r_{\rm h}$, most of the accreted stars would be bound to 
cause marginally eccentric TDEs. Otherwise, they would be unbound to cause marginally hyperbolic TDEs. 
We will have more detailed discussion about this speculation in the forthcoming paper \citep{zhong+18}.

%
%
%

%
%
%

%
%

\clearpage
%
\section{Discussion}
\label{sec:dis}
%

In a realistic intermediate-mass to supermassive system, the tidal disruption radius 
should be smaller than $\xi_{\rm acc}=10^{-5}$, which is the smallest normalized accretion 
radius among our simulations, if the accretion radius is equal to 
tidal disruption radius. Therefore, we extrapolate from the simulation data 
the orbital eccentricities at the realistic value of $\xi_{\rm acc}$ and 
the higher particle resolution by the linear least-square fitting method; $y=c{x}+d$, 
where the fitted values are plugged in $c$ and $d$, the mean value $e_{\rm mean}$ 
or standard deviation $e_{\rm std}$ of the orbital eccentricity are plugged in $y$, 
and the normalized accretion radius and the number of N-body particles are 
plugged in $x$, respectively. We define $e_{\rm mean}$ and $e_{\rm std}$ as follows: 
first, we divide the respective orbital eccentricities into some subsamples by a certain 
range of $\beta$, and then compute $e_{\rm mean}$ and $e_{\rm std}$ in each subsample. 
As one can see in the lower $\beta$ region around $\beta\sim1$, the different models 
show the significant variations, while in the higher $\beta$ region they take almost the 
same values. Therefore, the data points within the range of $1<\beta<1.2$ are used to 
calculate $e_{\rm mean}$ and $e_{\rm std}$.

\begin{figure}
\resizebox{\hsize}{!}{
\includegraphics{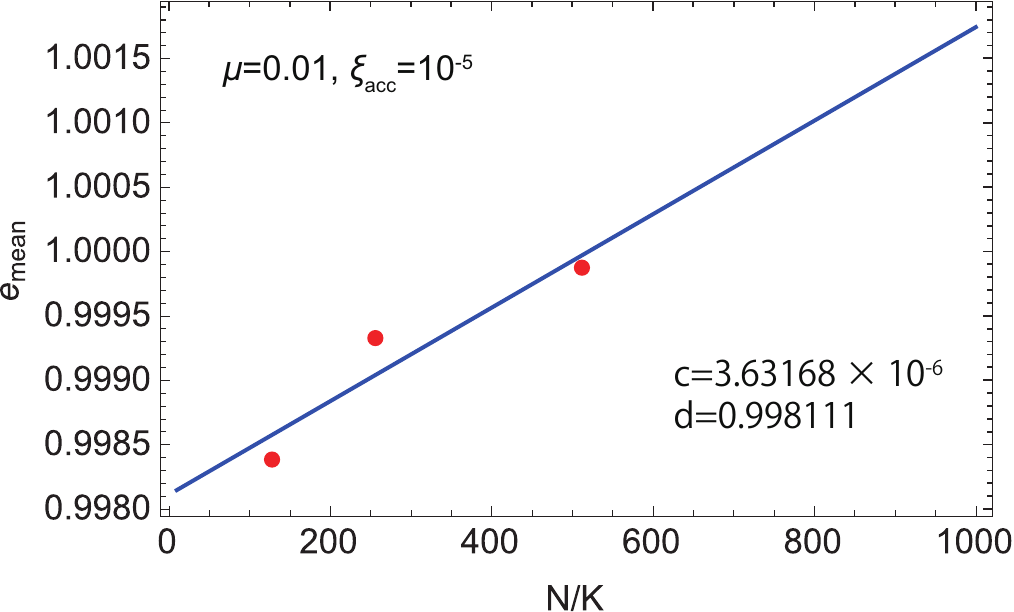}
\includegraphics{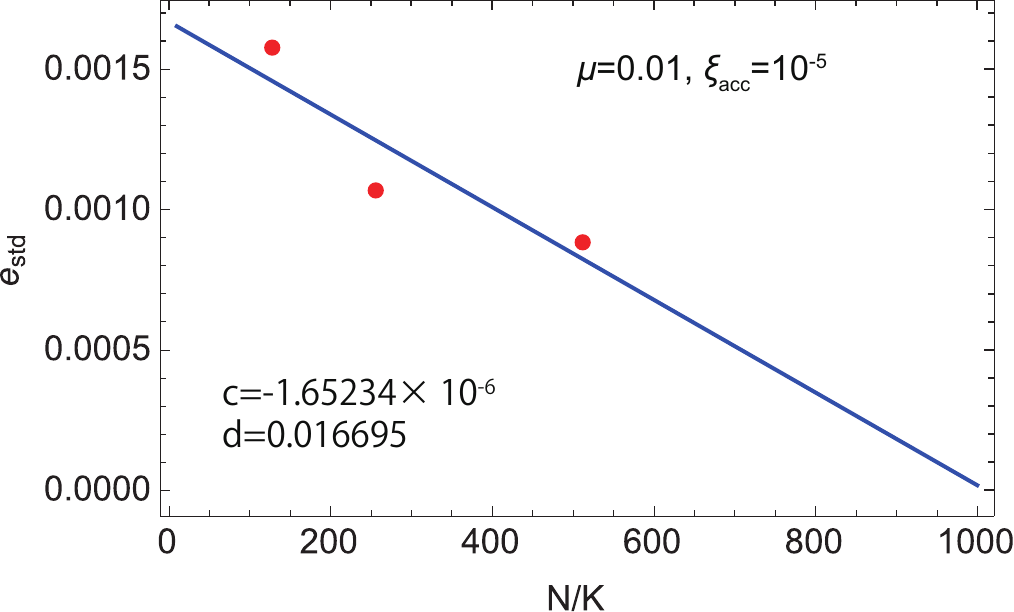}
}
\caption{
{
Dependence of the mean value and standard deviation 
of the orbital eccentricity on the number of the N-body 
particles for $\mu=0.01$ and $\xi_{\rm acc}=10^{-5}$. 
The red points shows them by numerical simulations, 
while the blue line shows the linear fitting by the 
least-square method; $y=c{x}+d$, where $y=e_{\rm mean}$ 
or $e_{\rm std}$ and $x=N/{\rm K}$, respectively.
}}
\label{fig:fit1}
\end{figure}

\begin{figure}
\resizebox{\hsize}{!}{
\includegraphics{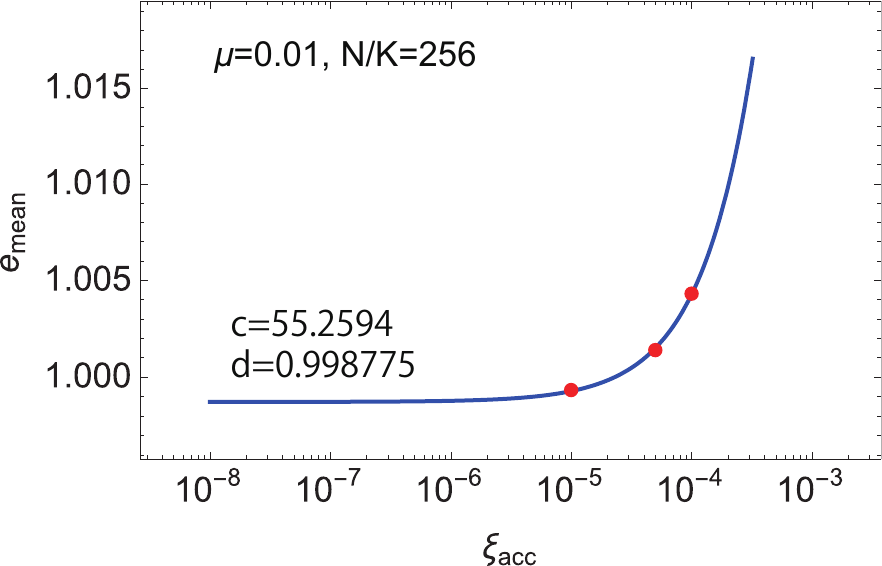}
\includegraphics{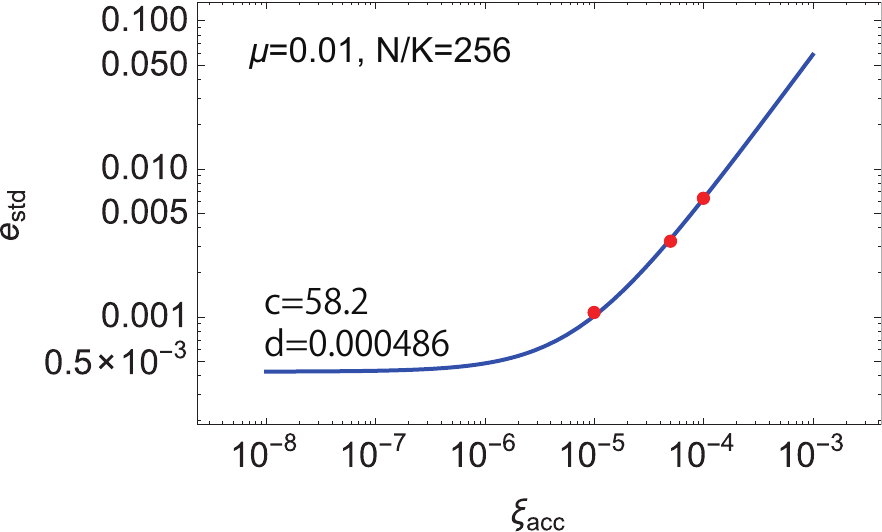}
}
\caption{
{
Dependence of the mean value and standard deviation 
of the orbital eccentricity on the normalized accretion 
radius for $\mu=0.01$ and $N/{\rm K}=256$. 
The red points shows them by numerical simulations, 
while the blue line shows the linear fitting by the 
least-square method; $y=c{x}+d$, where $y=e_{\rm mean}$ 
or $e_{\rm std}$ and $x=\xi_{\rm acc}$, respectively.
}}
\label{fig:fit2}
\end{figure}

Figure~\ref{fig:fit1} shows the dependence of the mean value and 
the standard deviation of the simulated orbital eccentricity on the 
number of the N-body particles for the fixed value of $\xi_{\rm acc}=10^{-5}$ 
and $\mu=0.01$ (Models 1, 3, and 5). The left and right panels are 
for $e_{\rm mean}$ and $e_{\rm std}$, respectively.
The eccentricity increases slightly 
with the higher mass resolution, whereas the 
standard deviation is smaller as the number of N-body particles increases.
Figure~\ref{fig:fit2} shows the dependence of the mean value and the standard 
deviation of the simulated orbital eccentricity on the normalized accretion radius 
for the fixed value of $N=256\,{\rm K}$ and $\mu=0.01$ (Models 3 and 4). 
Note that we used $\xi_{\rm acc}=5\times10^{-5}$ case to get the argument 
more reliable. Both the mean value and standard deviation of the orbital eccentricity 
decrease with the normalized accretion radius.
Overall, it is noted from the figures that the orbital eccentricity little deviates with 
the accretion radius and the number of N-body particles from the mean value.
This tendency can be adopted for the realistic extrapolated region, because 
the variation of the standard deviation is less than $0.1\%$ for the given variables.

Next, in order to discuss how realistic the extrapolated values are, we introduce 
\begin{eqnarray}
\rho_{\rm bh}\equiv{M_{\rm bh}}/r_{\rm t}^{3}
\label{eq:bhdens}
\end{eqnarray}
as the black hole density estimated at the tidal disruption radius. 
With the three dimensionless parameters we previously defined;
$\mu=M_{\rm bh}/M_{\rm c}$, $\xi_{\rm acc}=r_{\rm acc}/r_{\rm c}$, 
and $\zeta=r_{\rm t}/r_{\rm acc}$, the black hole density can be rewritten as
\begin{eqnarray}
\rho_{\rm bh}
=
\mu\left(\frac{\zeta}{\xi_{\rm acc}}\right)^{3}\rho_{\rm c},
\end{eqnarray}
where $\rho_{\rm c}\equiv{M}_{\rm c}/r_{\rm c}^{3}$ 
is defined as the mean stellar density of the cluster.
Substituting equation (\ref{eq:rt}) into (\ref{eq:bhdens}), 
the black hole density is equivalent with the mean star 
density; $\rho_{*}\equiv{m}_{*}/r_{*}^{3}$. Therefore, the 
normalized mean star density can be given in two separated 
forms as 
\begin{eqnarray}
\frac{\rho_{*}}{\rho_{\rm c}}=
\left\{ \begin{array}{ll}
\mu(\zeta/\xi_{\rm acc})^{3}=1.0\times10^{7}\,(0.01/\mu)^{-1}\,(1/\zeta)^{-3}\,
(10^{{-3}}/\xi_{\rm acc})^{3} \\
(m_{*}/r_{*}^{3})/\rho_{\rm c}=3.6\times10^{15}\,
(\rho_{\rm c}/10^{8}\,{\rm M_{\odot}\,pc^{-3}})^{-1}
(m_{*}/M_{\odot})(r_{*}/R_{\odot})^{-3}. \\
\end{array} \right.
\label{eq:rrhoc}
\end{eqnarray}
The upper equation shows the normalized mean star density obtained from our simulation 
parameters, where we adopt $\zeta=1$ which means that the accretion radius corresponds 
to the tidal disruption radius. In the lower equation, the normalized star density we estimate 
straightforwardly is constant for the accretion radius.

\begin{figure}
\resizebox{\hsize}{!}{
\includegraphics{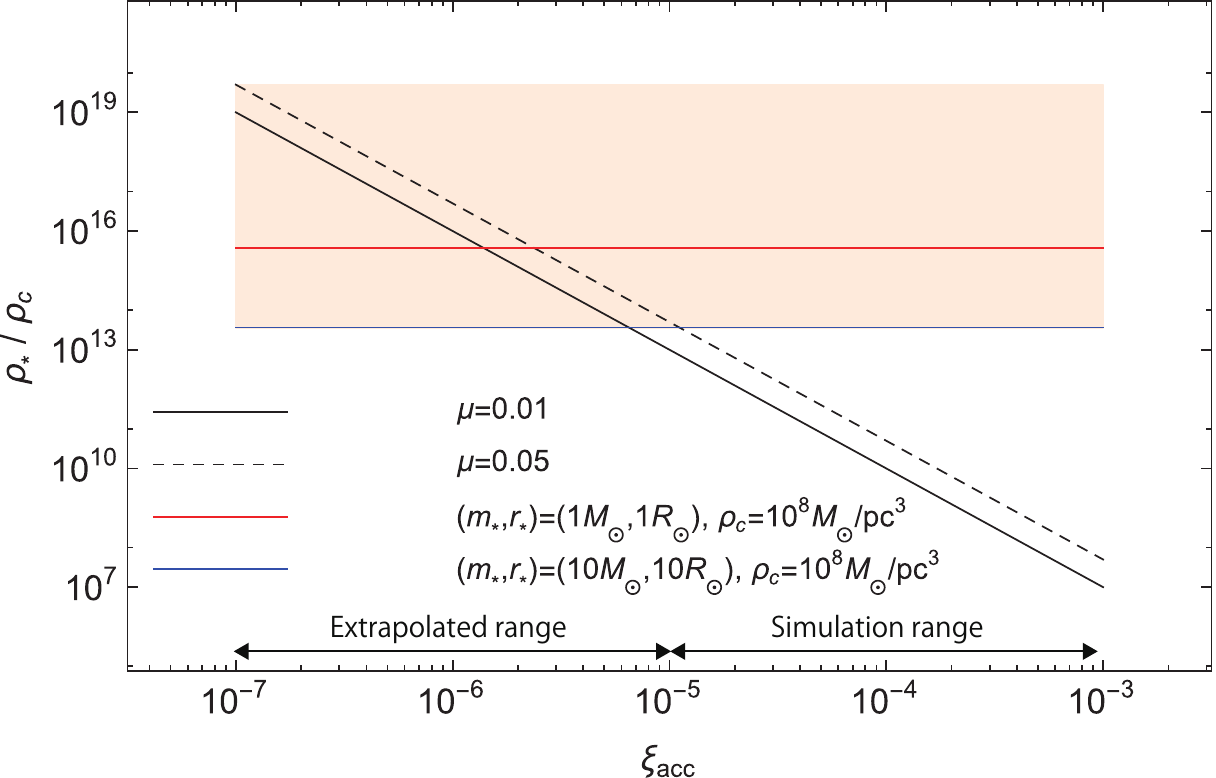}
}
\caption{
Dependence of the mean density of a star normalized by stellar density 
of a cluster on the accretion radius. The solid and dashed black lines are 
the normalized mean star density in the case of $\mu=0.01$ and $\mu=0.05$, 
respectively. The red and blue lines are those with $(m_{*},r_{*})=(1\,M_{\odot},1\,R_{\odot})$ 
and $(m_{*},r_{*})=(10\,M_{\odot},10\,R_{\odot})$, respectively, 
when $\rho_{\rm c}=10^{8}\,M_{\odot}\,{\rm pc^{-3}}$. 
The shaded area is the region where the stellar density of the cluster would be realistic. 
Our simulation models range from $\xi_{\rm acc}=10^{-3}$ to $10^{-5}$, whereas the 
extrapolated range is smaller than $\xi_{\rm acc}=10^{-5}$.
}
\label{fig:rhodist}
\end{figure}

Figure \ref{fig:rhodist} shows the dependence of the normalized mean star density 
on $\xi_{\rm acc}$. 
The solid and dashed black lines are $\rho_{*}/\rho_{\rm c}$, which is given by the upper 
part of equation (\ref{eq:rrhoc}), with $\mu=0.01$ and $\mu=0.05$, respectively. 
Assuming that $\rho_{\rm c}=10^{8}\,M_{\odot}\,{\rm pc}^{-3}$, the red and blue lines are 
$\rho_{*}/\rho_{\rm c}$, which is given by the lower part of equation (\ref{eq:rrhoc}), 
with $(m_{*},r_{*})=(1\,M_{\odot},1\,R_{\odot})$ and $(m_{*},r_{*})=(10\,M_{\odot},10\,R_{\odot})$, 
respectively. The shaded area is the region where the cluster density would be realistic. 
Our simulation models ranges from {$\xi_{\rm acc}=10^{-3}$ to $10^{-5}$}, whereas 
the extrapolated range is less than $\xi_{\rm acc}=10^{-5}$. From the figure, we note that the 
range of $\xi_{\rm acc}\lesssim10^{-5}$ should be realistic, if the averaged density of the 
realistic star cluster composing of mainly early type stars is equal to $10^{8}\,M_{\odot}\,{\rm pc}^{-3}$. 
This is independent of whether the cluster has a SMBH or IMBH. 

Let us discuss whether our extrapolation method is applicable to all the models we have done.
To resolve the transition from full to empty loss cone, as predicted by the loss cone theory 
\citep{fr76,2013CQGra..30x4005M}, in direct N-body simulations, $\xi_{\rm acc}$ has to be 
consistent with the limited resolution by the finite particle number in the model. Too large 
$\xi_{\rm acc}$ means all loss cones are too large and never completely empty (always 
$\theta_{\rm lc} > \theta_{\rm D}$), and too small one means that we are always in the pinhole 
regime where $\theta_{\rm D} > \theta_{\rm lc}$. For given particle number $N$ only a certain 
range of $\xi_{\rm acc}$ allows to resolve the correct full to empty loss cone transition at 
$\theta_{\rm lc} = \theta_{\rm D}$. \cite{zhong+14} confirmed that our simulations are consistent 
with the loss cone theory, if the normalized accretion radius is less than $\xi_{\rm acc}=10^{-4}$. 
Therefore, we applied our extrapolation method only for the simulation models with the normalized 
accretion radius less than $\xi_{\rm acc}=10^{-4}$. Model 15 should be excluded from the extrapolation 
argument noted above, although it produces a tiny but interesting possibility to cause both 
eccentric and hyperbolic TDEs, as shown in Table~2.

As seen in panel (b) of Figure~\ref{fig:class}, the critical eccentricities are also closer to unity 
as the ratio of the central black hole to stellar mass is larger. This tendency can be seen in 
Figure 4: the simulated critical eccentricities is close to unity as the black hole mass increases 
with time, although it is limited to the very narrow range of the mass ratio. 
In the forthcoming paper, we will examine the broader range of the mass ratio.

It is interesting to see which marginally eccentric or hyperbolic TDEs more preferably occur.
As discussed in Section 3.2, the source of the marginally eccentric TDEs is the stars falling 
to the black hole mainly from the critical radius inside the influence radius, whereas the source 
of the marginally hyperbolic TDEs is the stars falling to the black hole mainly from the critical 
radius outside the influence radius. Therefore, the ratio of $f_{\rm me}$ and $f_{\rm mp}$ 
should be determined by the location of the critical radius relative to the influence radius. 
This suggests that $f_{\rm me}/f_{\rm mp}$ is close to unity if the stars have a recessed 
distribution symmetrically around the radius, where the influence radius is accordingly equal 
to the critical radius. Models 5 and 10 correspond to this case. Whether this argument is 
robust would be confirmed by performing higher particle resolution N-body experiments 
with smaller accretion radius.

The deviation between some observed optical-UV TDEs light curves and 
the $t^{-5/3}$ decline rate is currently topics of debate (e.g. \citealt{sg+12}). 
Also, the soft X-ray TDE candidate represents the slightly different power 
law decay from $t^{-5/3}$ \citep{2013MNRAS.435.1904M}, although it 
looks corresponding to the $t^{-5/3}$ curve overall. 
Assuming that the observed luminosity is simply proportional to $t^{-n}$, 
we find 
\begin{equation}
n=\frac{2\alpha+5}{3}
\label{eq:nalpha}
\end{equation}
from our conjecture of the mass fallback rate given by equation (\ref{eq:mdot}). 
We note from Table~\ref{tbl:1} that the possible range of $n$ is 
$1<n\le5/3$ for the marginally eccentric and marginally hyperbolic TDEs. 
\cite{sg+12} discussed that the value of $n$ fitted to the decay 
of PS1-10jh was estimated to be $n=5/9$, $35/36$, and $12/15$ 
for the respective flaring phases. Because these indices are less than unity, 
our conjecture is not appropriate for PS1-10jh case. The other optical-UV 
TDE candidate, J0225-0432, represented that a best fit for the value of $n$ 
to the UV data gives $n\approx1.1$ \citep{2008ApJ...676..944G}. In this case, 
we cannot reject the possibility that J0225-0432 is a candidate for the marginally 
eccentric or marginally hyperbolic TDEs. This is also consistent with that the light curve 
of J0225-0432 should be shallower than the $t^{-5/3}$ profile by the internal 
structure of the star, as argued by \cite{lkp09}. 
If the star is partially disrupted, the range of $n$ can be from $2.2$ to $4$ 
because of the centrally condensed mass distribution, leading to the steeper 
mass fallback rate \citep{2013ApJ...767...25G}. This case is also beyond 
the scope of our conjecture. It suggests that the differential mass distribution 
should follow no simple power law of the specific energy. We need to rebuild 
the conjecture by taking account of the detailed internal structure of the star 
or the stellar debris.

%
Although these arguments seem independent of the semi-major axis and orbital 
eccentricity of the star approaching to the black hole, the difference between precisely 
parabolic and marginally eccentric/hyperbolic TDEs is shown in the magnitude 
of the mass fallback rate for a given value of $\alpha$. In addition, the value of 
$\alpha$ can depend on the semi-major axis and the orbital eccentricity 
as \cite{hsl13} implied. There is little known whether and how it can depend on them, 
and is no direct estimation of $\alpha$. Therefore, it is desired to examine the 
dependence of $\alpha$ on the given semi-major axis and orbital eccentricity in 
detail by the hydrodynamic simulations.

%
%

%
%

%
%

\pagebreak
%
\section{Conclusions}
\label{sec:con}
%

We have investigated the distribution of the orbital eccentricity of stars 
approaching the intermediate to supermassive black holes by N-body 
experiments. Since our N-body models do not reach a realistic resolution 
in particle number $N$ for galactic nuclei and consequently also cannot 
resolve the realistic value of the tidal disruption radius, we have used the 
method of scaling to extrapolate our results to the situation of a real galactic 
nucleus or nuclear stellar cluster. We have also found the condition to categorize 
the TDEs into the three types: eccentric, parabolic, and hyperbolic TDEs, from 
the viewpoints of the orbital eccentricity, $e$ and the semi-major axis of the originally 
approaching star, $a$. Based on the condition, we have analytically derived the mass 
fallback rates of respective TDEs. Our main conclusions are summarized as follows:
\begin{enumerate}
\item 
Parabolic TDEs are moreover divided into three subclasses: 
TDEs from stars on precisely parabolic orbits ($e=1$), 
marginally eccentric TDEs ($e_{\rm crit,1}\le{e}<1$), and 
marginally hyperbolic TDEs ($1<e\le{e}_{\rm crit,2}$).
While the mass fallback rate of marginally eccentric TDEs 
can be flatter and slightly higher than the standard fallback rate 
proportional to $t^{-5/3}$, it can be flatter and lower 
for marginally hyperbolic TDEs.
The detail is summarized in Table 1.
\item We find that there are two critical values of the orbital eccentricity: 
$e_{\rm crit,1}=1-2{q}^{-1/3}/\beta$ below which eccentric TDEs occur, 
and $e_{\rm crit,2}=1+2{q}^{-1/3}/\beta$ above which hyperbolic TDEs occur, 
where $q$ is the ratio of black hole to stellar mass and $\beta$ is the 
penetration factor. As the mass ratio is more extreme and the pericenter 
distance is closer to the Schwarzschild radius, these critical eccentricities 
are closer to 1. We confirm from our simulations that these critical 
eccentricities vary as the black hole grows.
\item Alternatively, there is a critical value of semi-major axis: 
$a_{\rm c}=50(q/10^6)^{1/3}r_{\rm t}$, where $r_{\rm t}$ is the tidal disruption radius. 
If $a\le{a}_{\rm c}$, then the eccentric and hyperbolic TDEs would occur. However, we 
confirm by N-body experiments that eccentric, precisely parabolic, and hyperbolic TDEs extremely rarely 
occur in a spherical stellar system with a single intermediate-mass to supermassive black 
hole. Instead, a substantial fraction of the stars causes marginally eccentric or marginally 
hyperbolic TDEs.
\end{enumerate}

%
%
\section*{Acknowledgments}
%
The authors thank the anonymous referee for fruitful comments and suggestions.
The authors also thank Nicholas~C.~Stone for his helpful comments and suggestions. 
This research has been supported by the Korea Astronomy and Space Science Institute (KASI) under 
the R\&D program supervised by the Ministry of Science, ICT and Future Planning and 
by Basic Science Research Program through the National Research Foundation of 
Korea (NRF) funded by the Ministry of Education (NRF-2017R1D1A1B03028580 K.H.). 
Data reductions were performed by using a high performance computing cluster at the KASI. 
Authors acknowledge support by the Chinese Academy of Sciences (CAS) through the Silk 
Road Project at NAOC, through the CAS Visiting Professorship for Senior International Scientists, 
Grant Number 2009S1-5 (R.S.), and through the 'Qianren' special foreign experts programme of 
China. S.Z., S.L., and R.S. have been partially supported by National Natural Science Foundation 
of China (NSFC 11603067 S.Z., 11303039 S.L., and 11673032 R.S.). P.B. acknowledges the special 
support by the NASU under the Main Astronomical Observatory GRID/GPU computing cluster project. 
The special GPU accelerated supercomputer laohu at the Center of Information and Computing 
at National Astronomical Observatories, CAS, funded by Ministry of Finance of People's Republic 
of China under the grant ZDYZ2008-2 has been used for computer simulations. S.L. P.B., and R.S. 
acknowledge also the Strategic Priority Research Program (PilotB) ''Multi-waveband Gravitational 
Wave Universe'' of the CAS (No. XDB23040100).

%

\end{document}